\documentclass[12pt]{article}
\usepackage{amssymb}
\usepackage{amsfonts}
\usepackage{graphics}
\usepackage{amsmath}
\usepackage{color}
\usepackage{breqn}
\usepackage{graphicx}
\usepackage{caption}
\usepackage{subcaption}
\usepackage{pdflscape}
\usepackage[official]{eurosym}
\usepackage{array}
\usepackage[hyphens]{url}
\usepackage[flushleft]{threeparttable} 
\usepackage[utf8]{inputenc}
\usepackage{floatrow}
\usepackage{bm}
\usepackage{bbm}
\usepackage{booktabs}
\newcommand{\norm}[1]{\left\lVert#1\right\rVert}
\usepackage{makecell}
\usepackage{pdflscape}
\usepackage{comment}
\usepackage{sectsty}
\usepackage{setspace}
\usepackage{appendix}
\usepackage{resizegather}
\usepackage{hyperref}
\usepackage[round]{natbib}
\usepackage{breakcites}

\newcommand{\plim}{\operatorname*{plim}}

\DeclareMathOperator{\vect}{vec}

\newtheorem{theorem}{Theorem}
\newcounter{coro}
\newtheorem{corollary}[coro]{Corollary}
\newcounter{def}
\newcounter{prop}
\newcounter{lem}
\newtheorem{lemma}[lem]{Lemma}

\newcounter{probl}
\newcounter{ex}
\newenvironment{example}
{\refstepcounter{ex}\bigskip\noindent\textbf{Example~\arabic{ex}\ \
}}{\
	\rule{0.5em}{0.5em}\medskip}
\newcounter{assum}
\newtheorem{assumption}[assum]{Assumption}
\newcounter{rem}
\newenvironment{remark}
{\refstepcounter{rem}\bigskip\noindent\textbf{Remark~\arabic{rem}\ \
}}{\
	\rule{0.5em}{0.5em}\medskip}

\refstepcounter{figure}

\refstepcounter{table}

\newcommand{\sfrac}[2]{\ensuremath\raisebox{1.5pt}{\footnotesize
		$#1$}\kern-1pt/\kern-1pt
	\raisebox{-2pt}{\footnotesize $#2$}}
\def\leaderfill{\leaders\hbox to 1em{\hss.\hss}\hfill}
\def\smallskip{\vskip\smallskipamount}
\def\medskip{\vskip\medskipamount}
\def\bigskip{\vskip\bigskipamount}
\begin{document}
	
	\title{Cluster-Robust Standard Errors for Linear Regression Models with Many Controls\footnote{I am grateful to Martin Weidner for his help and support since the start of this project. I would also like to thank Ivan Canay, Andrew Chesher, Whitney Newey, Valentin Verdier and participants at the Bristol Econometric Study Group for helpful comments and discussions. \newline$^\dagger$Department of Economics, University College London. E-mail: \href{mailto:riccardo.d'adamo.15@ucl.ac.uk}{\texttt{riccardo.d'adamo.15@ucl.ac.uk}}}}
	\author{Riccardo D'Adamo$^\dagger$}
	
	\maketitle

\begin{abstract}
	
It is common practice in empirical work to employ cluster-robust standard errors when using the linear regression model to estimate some structural/causal effect of interest. Researchers also often include a large set of regressors in their model specification in order to control for observed and unobserved confounders. In this paper we develop inference methods for linear regression models with many controls and clustering. We show that inference based on the usual cluster-robust standard errors by \citet*{liangzeger} is invalid in general when the number of controls is a non-vanishing fraction of the sample size. We then propose a new clustered standard errors formula that is robust to the inclusion of many controls and allows to carry out valid inference in a variety of high-dimensional linear regression models, including fixed effects panel data models and the semiparametric partially linear model. Monte Carlo evidence supports our theoretical results and shows that our proposed variance estimator performs well in finite samples. The proposed method is also illustrated with an empirical application that re-visits \citeauthor*{dl2001}'s (\citeyear{dl2001}) study of the impact of abortion on crime.

\end{abstract}
\let\markeverypar\everypar
\newtoks\everypar
\everypar\markeverypar
\markeverypar{\the\everypar\looseness=-2\relax}
\thispagestyle{empty}
\parskip=3.pt	
\newpage
\baselineskip=20pt

\section{Introduction}\label{sec:Introduction}
\setcounter{page}{1}

It is common practice in empirical work to use standard errors and associated confidence intervals that are robust to heteroskedasticity and/or various forms of dependence. 
In particular, since \citet*{moulton} highlighted the importance of accounting for dependence arising in data with a group structure, researchers often assume that the data are clustered at some economically relevant level, e.g. by individual unit or geographical location. 

The justification for this type of inference procedures is asymptotic, in the sense that their validity relies on the assumption that the sample size is large relative to the number of parameters in the model. In small samples two issues arise: (i) confidence intervals based on the usual Gaussian approximation become invalid, (ii) robust standard errors are biased. A variety of methods that address these issues in the context of the linear regression model have been proposed in the literature. However, these usually alleviate but do not entirely solve the problem and are not always appealing given their ad hoc nature (see \citealp{imbenskolesar}, for a discussion).

Furthermore, while modern datasets usually include a large number of observations, the assumption that the number of parameters in the estimated model is negligible relative to the sample size can still be unattractive, even when the researcher's goal is to conduct inference on a small set of parameters. For example, in many important applications of the linear regression model, the object of interest is $\bm{\beta}$ in a model of the form

 \begin{equation}\label{model}
y_{i,n}=\bm{\beta}'\mathbf{x}_{i,n} + \bm{\gamma}_n'\mathbf{w}_{i,n} + u_{i,n},\qquad i=1,\dots, n,
\end{equation}
where $y_{i,n}$ is a scalar outcome variable, $\mathbf{x}_{i,n}$ is a $d\times1$ vector of regressors of fixed dimension, $\mathbf{w}_{i,n}$ is a vector of covariates of possibly ``large" dimension $K_n$, and $u_{i,n}$ is an unobserved scalar error term. In many applications of this model, the assumption that $K_n/n\to0$ is unpalatable or even violated, as researchers often include a large set of covariates in $\mathbf{w}_{i,n}$ in order to control for observed and unobserved confounders (see discussion below).

Motivated by the above observations, this paper develops inference theory for linear regression models with many controls and clustering. In particular, we first show that the usual cluster-robust standard errors by \citet*{liangzeger} are inconsistent in general when $K_n/n\nrightarrow 0$. We then propose a new clustered standard error formula that allows to carry out valid inference on $\bm{\beta}$ under asymptotics in which $K_n$ is allowed (but not required) to grow as fast as the sample size. 

The findings of this paper contribute to the long-established literature initiated by \citet*{whitebook} dealing with cluster-robust inference in a variety of models, a recent review of which is given by \citet*{cameronmiller}; see, e.g., \citet*{arellano1987}, \citet*{bellmccaffrey}, \citet*{hansen2007}, \citet{cameron2008}, \citet*{ibragimovmuller}, \citet*{pustejovskytilton} and \citet{canaycluster2008}. In particular, our analysis is related to a literature, reviewed in \citet*{imbenskolesar}, in which bias-reduction modifications of standard errors and particular distributional approximations are proposed with the aim of improving the performance of cluster-robust inference procedures in small samples. We contribute to this literature by studying a new general class of cluster-robust variance estimators that allows to fully correct such ``small-sample" bias, while also exploiting the particular structure of the model in (\ref{model}) to circumvent the need for non-Gaussian distributional approximations. 

This paper also adds to a sizeable body of literature that deals with inference procedures in models that involve the estimation of many incidental parameters; see, e.g., \citet*{angristhahn2004}, \citet*{hahnewey}, \citet*{stockwatson2008}, \citet{belloni2014rev}, \citeauthor*{cjn} (\citeyear{cjn}, \citeyear{cattaneo2018partially}), \citet*{verdier}, and references therein. In particular, our findings can be seen as generalising those of \citet*{cjn}, CJN hereafter, who establish asymptotic normality of the OLS estimator of $\bm{\beta}$ in (\ref{model}) when $K_n/n\nrightarrow0$, and provide inference methods under such asymptotics when the errors are independent and heteroskedastic.

The results in this paper were derived independently of \citet*{lithesis}, who tackles the problem of cluster-robust variance estimation for the full set of coefficients of a generic high-dimensional linear model and obtains a similar estimator to ours. However, his results are not directly applicable to inference and are silent about the extent to which sufficient conditions for consistent variance estimation restrict the underlying data generating process of the regressors.

The rest of this paper is organized as follows. Section 2 introduces the framework of the paper and illustrates its relevance using three leading examples. Section 3 discusses our assumptions. Section 4 presents our main theoretical results. Section 5 reports the findings of a Monte Carlo study. Section 6 presents an empirical illustration. Section 7 briefly concludes. Proofs and extensions of the results are given in the Appendix.

\section{Framework and Motivation}
The main object of interest in our analysis is $\bm{\beta}$ in (\ref{model}), on which we would like to carry out inference while treating the high-dimensional $\mathbf{w}_{i,n}$ as nuisance covariates. A natural choice of estimator for $\bm{\beta}$ is the OLS estimator, which can be written as
\begin{equation}\label{partialledout}
\bm{\hat{\beta}}=(\sum_{i=1}^{n}\mathbf{\hat{v}}_{i,n}\mathbf{\hat{v}}_{i,n}')^{-1}(\sum_{i=1}^{n}\mathbf{\hat{v}}_{i,n}y_{i,n}), \qquad \mathbf{\hat{v}}_{i,n}=\sum_{j=1}^{n}M_{ij,n}\mathbf{x}_{j,n}, 
\end{equation}
where $M_{ij,n}=\mathbbm{1}\{i=j\}-\mathbf{w}_{i,n}'(\sum_{k=1}^{n}\mathbf{w}_{k,n}\mathbf{w}_{k,n}')^{-1}\mathbf{w}_{j,n}$  is the $(i,j)$ entry of the symmetric and idempotent annihilator matrix $\mathbf{M}_n$, with $\mathbbm{1}\{\cdot\}$ denoting the indicator function.
Defining $\bm{\hat{\Gamma}}_n=\sum_{i=1}^n\mathbf{\hat{v}}_{i,n}\mathbf{\hat{v}}_{i,n}'/n$ and $\bm{\Sigma}_n$ the (conditional) variance of $\sum_{i=1}^n\mathbf{\hat{v}}_{i,n}u_{i,n}/\sqrt{n}$, it is well-known that, when $n\to\infty$ and $K_n$ is fixed, the asymptotic distribution of $\bm{\hat{\beta}}_n$ is
\begin{equation}
\bm{\Omega}_n^{-1/2}\sqrt[]{n}(\bm{\hat{\beta}}_n-\bm{\beta})\overset{d}{\to}\mathcal{N}(0,\mathbf{I}_d), \qquad \bm{\Omega}_n=\bm{\hat{\Gamma}}^{-1}_n\bm{\Sigma}_n\bm{\hat{\Gamma}}^{-1}_n.
\end{equation}
When the errors are assumed to be correlated only within $G_n$ clusters of bounded size, $\bm{\Sigma}_n$ can be estimated consistently with the popular cluster-robust variance estimator by Liang and Zeger (1986, LZ hereafter):
\begin{equation}
\bm{\hat{\Sigma}}_n^{\texttt{LZ}}=\frac{1}{n}\sum_{g=1}^{G_n}\sum_{i,j\in\mathcal{T}_{g,n}}\mathbf{\hat{v}}_{i,n}\mathbf{\hat{v}}'_{j,n}\hat{u}_{i,n}\hat{u}_{j,n}, \qquad \hat{u}_{i,n}=\sum_{j=1}^nM_{ij,n}(y_{j,n}-\bm{\hat{\beta}}_n'\mathbf{x}_{j,n}),
\end{equation}
where $\mathcal{T}_{g,n}$ denotes the subset of observations contained in cluster $g$ and $\{\mathcal{T}_{g,n}:1\leq g\leq G_n\}$ is a partition of the data. As a result, asymptotically valid inference can be carried out using the usual testing procedures based on the distributional approximation $\bm{\hat{\beta}}_n\stackrel{a}{\sim}\mathcal{N}(\bm{\beta},\bm{\hat{\Gamma}}_n^{-1}\bm{\hat{\Sigma}}_n^{\texttt{LZ}}\bm{\hat{\Gamma}}_n^{-1}/n)$.

The objective of this paper is to establish cluster-robust inference procedures for $\bm{\beta}$ under asymptotics in which $K_n/n\nrightarrow 0$. Allowing the dimension of the nuisance covariates $K_n$ to grow at the same rate as the sample size $n$ enables us to cover many relevant applications of the general model in (\ref{model}).
\begin{example}{\textbf{Linear regression model with increasing dimension}}\\
This leading example takes (\ref{model}) as the data generating process, in which $\mathbf{w}_{i,n}$ contains many observable individual characteristics and their nonlinear transformations, dummy variables for many categories such as age group, cohort, geographic location etc. and their interactions with the former. The inclusion of many covariates is motivated in practice by the assumption that the variable of interest $\mathbf{x}_{i,n}$ can be taken as exogenous after controlling for $\mathbf{w}_{i,n}$.  
Although the study of linear regression models with growing dimension has a long tradition in statistics (see, e.g., Huber, 1973, and Mammen, 1993), until recently inference results were exiguous and limited to the case in which the number of regressors in the model is at least a vanishing fraction of the sample size. \citet{cjn} exploit the separability of model (\ref{model}) to develop valid inference procedures for $\bm{\beta}$ when $K_n/n\nrightarrow 0$, but their theory only covers the case of homoskedastic and heteroskedastic errors. \citet*{limuller} develop cluster-robust inference theory in this setting for a scalar $\beta$, i.e. $d=1$; their results allow for $K_n\propto n$ but rely on a strong restriction on $\sum_{i=1}^{n}(\bm{\gamma}_n'\mathbf{w}_{i,n})^2$, which limits the amount of sample variation of $y_i$ that can be induced by the high-dimensional controls $\mathbf{w}_{i,n}$. \citet{belloni2014rev} instead propose an estimation procedure for $\bm{\beta}$ based on LASSO double-selection and provide inference theory for the case of i.n.i.d. data. While their method can accomodate $K_n\gg n$, it relies on the assumption that the effect of confounders can be controlled for by a small subset of the variables in $\mathbf{w}_{i,n}$ up to some small approximation error (``approximate sparsity").
\end{example}
\begin{example}{\textbf{Semiparametric Partially Linear Model}}\label{example2}\\
Researchers often assume that data are generated by the model
\begin{equation}\label{partially_linear_model}
y_i=\bm{\beta}'\mathbf{x}_i +g(\mathbf{z}_{i})+\varepsilon_i, \qquad i=1,\dots, n, \end{equation}
where both $\mathbf{x}_{i}$ and $\mathbf{z}_{i}$ have fixed dimension, but the function $g(\cdot)$ is unknown. The partially linear model is a long-standing area of interest in econometrics (see, e.g., Heckman, 1986, and Robinson, 1988). Estimation of this semiparametric model is often carried out via series-based methods, in which the researcher assumes that the function $g(\cdot)$ can be closely approximated using the polynomial functions $\mathbf{p}_n(\mathbf{z})=(p^1(\mathbf{z}), \dots, p^{K_n}(\mathbf{z}))'$, so that $g(\mathbf{z}_i)\approx\bm{\gamma}_n'\mathbf{p}_n(\mathbf{z}_i)$ for some $\bm{\gamma}_n$. The series estimator for $\bm{\beta}$ is the OLS estimator as defined in (\ref{partialledout}), where $\mathbf{w}_{i,n}=\mathbf{p}_n(\mathbf{z}_i).$ When the underlying function $g(\cdot)$ is not sufficiently smooth and/or the dimension of $\mathbf{z}_i$ is relatively large, the inclusion of many polynomial terms might be required, resulting in $K_n$ being non-negligible relative to $n$. \citet{cattaneo2018partially} are the first to consider asymptotics in which $K_n/n\nrightarrow0$ in this setting. They establish asymptotic normality for $\bm{\hat{\beta}}_n$ and valid inference procedures under homoskedasticity and, in their subsequent paper, heteroskedasticity (CJN, 2018a). However, no results are available for the case of clustering.
\end{example}
\begin{example}{\textbf{Multi-way fixed effects panel data models}}\\
Panel data models that use fixed effects are often used in order to control for unobserved heterogeneity, such as the one-way fixed effects panel data regression model
\begin{equation}\label{oneway}
Y_{it}=\bm{\beta}'\mathbf{X}_{it} +\alpha_ i + U_{it}, \qquad i=1,\dots, N, \qquad t=1,\dots, T,\end{equation}
where $\alpha_i$ is a scalar individual effect, $\mathbf{X}_{it}$ is a vector of regressors and $U_{it}$ is a scalar error term. This model can be mapped into our baseline specification in (\ref{model}) by setting $n=NT$, $y_{(i-1)T+t,n}=Y_{it}$, $\mathbf{x}_{(i-1)T+t,n}=\mathbf{X}_{it}$, $u_{(i-1)T+t,n}=U_{it}$, $\bm{\gamma}_n=(\alpha_1, \dots, \alpha_N)$ and $\mathbf{w}_{(i-1)T+t,n}$ equal to the $i$-th unit vector of dimension $N$. It follows that $K_n=N$ and $K_n/n=1/T$, which motivates the asymptotics of this paper under $N\to\infty$ and T fixed. For this case, \citet*{arellano1987} shows that LZ's variance estimator (1986) is consistent when errors are clustered at the individual level. For the same setting, \citet*{stockwatson2008} propose a cluster-robust estimator for the variance with additional zero restrictions on the conditional autocovariances of the errors  within entities, e.g. when an MA(q) structure is imposed on $U_{it}$. 

In many empirical settings, researchers want to control for multiple terms of unobserved heterogeneity. In the analysis of student/teacher or worker/firm matched data, for example, two-way fixed effects models are commonly used, taking the form
\begin{equation}\label{twoway}
Y_{it}=\bm{\beta}'\mathbf{X}_{it} +\alpha_ i + e_{d_{it}}+ U_{it}, \qquad i=1,\dots, N, \qquad t=1,\dots, T,
\end{equation}
where $e_{d_{it}}$ are unobserved factors common to all observations sharing the same value of the indexing variable $d_{it}\in\{1,\dots, N_d\}$, so that $\mathbf{w}_{(i-1)T+t,n}$ is now a $N+N_d$ vector selecting the relevant fixed effects from $\bm{\gamma}_n=(\alpha_1, \dots, \alpha_N, e_1, \dots, e_{N_d})$. When T is fixed and only few observations are assigned to each value of $d_{it}$ (i.e. data are sparsely matched), then the number of fixed effects grows proportionally to the sample size and $K_n\propto n$. \citet*{verdier} considers cluster-robust inference under these asymptotics for a new estimation procedure for $\bm{\beta}$ that accomodates instrumental variables but is generally less efficient than OLS. He also proposes a variance estimator that can be seen as a generalisation of the one proposed in this paper to multi-way cluster dependence. 
\end{example}

To simplify exposition, we present our inference theory for linear regression models with many controls for the case of strictly exogenous regressors. While all the results of this paper can be well-understood for this special case, their generalisation to (potential) misspecification bias in the model is straightforward and is provided in the Appendix.

\section{Assumptions} 
In this section we present a set of assumptions for the special case of strict exogeneity of the regressors. A more general set of assumptions that allows for misspecification bias is given in the Appendix.

Suppose that $\{(y_{i,n},\mathbf{x}_{i,n}',\mathbf{w}_{i,n}') : 1 \leq i \leq n\}$ is generated by (\ref{model}) and set $\mathcal{X}_{n}=(\mathbf{x}_{1,n},\dots,\mathbf{x}_{n,n})$ and $\mathcal{W}_{n}=(\mathbf{w}_{1,n},\dots,\mathbf{w}_{n,n})$. We define the following quantities:
\begin{equation*}
\begin{split}
&\chi_n=\frac{1}{n}\sum_{i=1}^{n}\mathbb{E}[\|\mathbf{Q}_{i,n}\|^2], \qquad \mathbf{Q}_{i,n}=\mathbb{E}[\mathbf{v}_{i,n}|\mathcal{W}_n],\\
&\bm{\hat{\Gamma}}_n=\sum_{i=1}^n\mathbf{\hat{v}}_{i,n}\mathbf{\hat{v}}_{i,n}'/n, \qquad \bm{\Sigma}_n=\mathbb{V}[\frac{1}{\sqrt{n}}\sum_{i=1}^{n}\mathbf{\hat{v}}_{i,n}u_{i,n}|\mathcal{X}_n,\mathcal{W}_n],
\end{split}
\end{equation*}
where $\mathbf{v}_{i,n}=\mathbf{x}_{i,n}-(\sum_{j=1}^{n}\mathbb{E}[\mathbf{x}_{j,n}\mathbf{w}_{j,n}'])(\sum_{j=1}^{n}\mathbb{E}[\mathbf{w}_{j,n}\mathbf{w}_{j,n}'])^{-1}\mathbf{w}_{i,n}$ is the population counterpart of $\mathbf{\hat{v}}_{i,n}$. Also, letting $\lambda_{\text{min}}(\cdot)$ denote the minimum eigenvalue of its argument, define
\begin{equation*}
\mathcal{C}_n=\max_{1\leq i \leq n}{\{\mathbb{E}[u_{i,n}^4|\mathcal{X}_n,\mathcal{W}_n]+\mathbb{E}[\|\mathbf{V}_{i,n}\|^4 |\mathcal{W}_n]+1/\mathbb{E}[u_{i,n}^2|\mathcal{X}_n,\mathcal{W}_n]}\}+1/\lambda_{\text{min}}(\mathbb{E}[\bm{\tilde{\Gamma}}_n|\mathcal{W}_n])
\end{equation*}
where $\mathbf{V}_{i,n}=\mathbf{x}_{i,n}-\mathbb{E}[\mathbf{x}_{i,n}|\mathcal{W}_n]$, $\mathbf{\tilde{\Gamma}}_n=\sum_{i=1}^{n}\mathbf{\tilde{V}}_{i,n}\mathbf{\tilde{V}}_{i,n}'/n$ and $\mathbf{\tilde{V}}_{i,n}=\sum_{j=1}^{n}M_{ij,n}\mathbf{V}_{i,n}$.\\
\par We impose the following three assumptions:
\begin{assumption}{}
$\max_{1\leq g\leq G_n}\#\mathcal{T}_{g,n}=O(1)$, where $\#\mathcal{T}_{g,n}$ is the cardinality of $\mathcal{T}_{g,n}$ and where $\{\mathcal{T}_{g,n}:1\leq g\leq G_n\}$ is a partition of \{1,\dots,n\} such that $\{(u_{i,n},\mathbf{x}'_{i,n}):i\in\mathcal{T}_{g,n}\}$ are independent over g conditional on $\mathcal{W}_{n}$. 
\end{assumption}
\begin{assumption}{}
 $\mathbb{P}[\lambda_{\text{min}}(\sum_{i=1}^{n}\mathbf{w}_{i,n}\mathbf{w}_{i,n}')>0]\rightarrow 1$, $\limsup_{n\rightarrow\infty}K_n/n < 1$, $\mathcal{C}_n=O_p(1)$ and $\bm{\Sigma}_n^{-1}=O_p(1)$
\end{assumption}
\begin{assumption}
$\mathbb{E}[u_{i,n}|\mathcal{X}_n,\mathcal{W}_n]=0\quad \forall i,n$, $\chi_n=O(1)$,and $\max_{1\leq i \leq n}\|\mathbf{\hat{v}}_{i,n}\|/\sqrt{n}=o_p(1)$.
\end{assumption}
Assumption 1 defines the sampling structure, in which we allow for arbitrary dependence within clusters of finite but possibly heterogenous size for both the regressors and the errors. In terms of clustering structure, the resulting asymptotics are the same as the usual ones of \citet*{whitebook} and \citet{liangzeger} in which $n, G_n\to\infty$ and $G_n\propto n$. We expect that the results of this paper would generalize to asymptotics where cluster sizes are allowed to diverge with $n$ and $G_n$, as considered in \citet*{hansen2007} and \citet*{hansenlee}. It is likely that such extension would require imposing more restrictive conditions on the regression design and the distributional properties of the errors, e.g. stationarity and/or mixing, and we leave it to future work.

Assumption 2 allows for asymptotics where $K_n/n\nrightarrow 0$, while imposing standard restrictions on the regression design and some bounds on the (conditional) higher-order moments of the structural residuals $u_{i,n}$ and $\mathbf{V}_{i,n}$.

The condition on $\chi_n$ in Assumption 3 is a requirement on the quality of the linear approximation for the conditional expectation $\mathbb{E}[\mathbf{x}_{i,n}|\mathcal{W}_n]$. The high-level condition $\max_{1\leq i \leq n}\|\mathbf{\hat{v}}_{i,n}\|/\sqrt{n}=o_p(1)$ also places restrictions on the relationship between $\mathbf{x}_{i,n}$ and $\mathbf{w}_{i,n}$ and has a central importance in establishing asymptotic normality of the OLS estimator for $\bm{\beta}$ and consistency of our proposed variance estimator. \citet{cjn} show that this restriction holds under mild moment conditions when either (i) $K_n/n\to0$, or (ii) $\chi_n=o(1)$ or (iii) $\max_{1\leq i \leq n}\sum_{i=1}^n\mathbbm{1}\{M_{ij,n}\neq0\}=o_p(n^{1/3})$. While condition (i) is not the case of primary interest of this paper, (ii) and (iii) accomodate $K_n/n\nrightarrow 0$ and can be used to verify the high-level condition $\max_{1\leq i \leq n}\|\mathbf{\hat{v}}_{i,n}\|/\sqrt{n}=o_p(1)$ when $\mathbf{w}_{i,n}$ can be interpreted as approximating functions, dummy/discrete variables or fixed effects. See \citet{cjn} for details.
\begin{remark}
In the general formulation provided in Appendix, the assumptions we consider are analogous to those in \citet{cjn} but we also allow for clustered dependence in the errors. The set of restrictions imposed by this framework allows to cover the three leading examples presented in the previous section. A detailed discussion of conditions that satisfy the assumptions in those particular models is provided in \citet{cjn} and their Supplemental Appendix.
\end{remark}

\section{Main Results}
This section presents our main theoretical results for inference in linear regression models with many controls and clustering under the set of simplified assumptions presented in Section 3. Proofs of the theorems and other auxiliary results are given in the Appendix for the general case that allows for misspecification bias in the model.

Our first result extends the asymptotic normality result for $\hat{\bm{\beta}}$ previously derived by \citet{cjn} to the case of clustering.
\begin{theorem}{}
 Suppose Assumptions 1-3  hold. Then,
\begin{equation*}
\bm{\Omega}_n^{-1/2}\sqrt[]{n}(\bm{\hat{\beta}}_n-\bm{\beta})\overset{d}{\to}\mathcal{N}(0,\mathbf{I}_d), \qquad \bm{\Omega}_n=\bm{\hat{\Gamma}}^{-1}_n\bm{\Sigma}_n\bm{\hat{\Gamma}}^{-1}_n,
\end{equation*}
where $\bm{\Sigma}_n=\frac{1}{n}\sum_{g=1}^{G_n}\sum_{i,j\in\mathcal{T}_{g,n}}\mathbf{\hat{v}}_{i,n}\mathbf{\hat{v}}'_{j,n}\mathbb{E}[u_{i,n}u_{j,n}|\mathcal{X}_n,\mathcal{W}_n]$.
\end{theorem}

Theorem 1 implies that the asymptotic distribution of $\hat{\bm{\beta}}$ under $K_n/n\nrightarrow0$ resembles the standard one
obtainable under fixed-$K_n$.\footnote{From Assumptions 1-3 it also follows that $\bm{\hat{\Omega}}_n=O_p(1)$, implying that $\bm{\hat{\beta}}_n$ is $\sqrt{n}$-consistent.} As a result, confidence intervals can be constructed using the usual Gaussian approximation and the problem of conducting valid inference reduces to finding a consistent estimator for $\mathbf{\Sigma}_n$ under our asymptotics of interest. 

For our discussion of variance estimation, we introduce a new class of estimators. 
Let $\bm{\Omega}_{u,n}=\mathbb{E}[\mathbf{u}_n\mathbf{u}_n'|\mathcal{X}_n,\mathcal{W}_n]$ be the (conditional) variance-covariance matrix of the errors $\mathbf{u}_{n}=(u_{1,n}, \dots, u_{n,n})'$ and $L_n=\sum_{g=1}^{G_n}(\#\mathcal{T}_{g,n})^2$ the number of non-zero elements contained in it.
We define a general class of cluster-robust estimators for $\bm{\Sigma}_n$ of the form
\begin{equation}\label{clusterclass}
\bm{\hat{\Sigma}}_n^{}(\bm{\kappa}_n)=\frac{1}{n}\sum_{g_1=1}^{G_n}\sum_{g_2=1}^{G_n}\sum_{i_1,j_1\in\mathcal{T}_{g_1,n}}\sum_{i_2,j_2\in\mathcal{T}_{g_2,n}}\kappa_{g_1, g_2, i_1,j_1,i_2,j_2,n}\mathbf{\hat{v}}_{i_1,n}\mathbf{\hat{v}}'_{j_1,n}\hat{u}_{i_2,n}\hat{u}_{j_2,n},
\end{equation}
where $\kappa_{g_1,g_2,i_1,j_1,i_2.j_2,n}$ is an entry of the $L_n\times L_n$ symmetric matrix $\bm{\kappa}_{n}=\bm{\kappa}_{n}(\mathbf{w}_{1,n}, \dots, \mathbf{w}_{1,n})$.\footnote{In particular, $\kappa_{g_1,g_2,i_1,j_1,i_2.j_2,n}$ corresponds to the $(h(g_1,i_1,j_1), h(g_2,i_2,j_2))$ entry of $\bm{\kappa}_{n}$, where $h(g,i,j)=[\sum_{k=0}^{(g-1)}(\#\mathcal{T}_{k,n})^2+(\#\mathcal{T}_{g,n})(i-1) +j]$ and we adopt the convention that $\#\mathcal{T}_{0,n}=0$.}
Notice that by setting $\bm{\kappa}_n=\mathbf{I}_{L_n}$ one obtains the usual cluster-robust estimator by \citet{liangzeger}:
\begin{equation*}
\bm{\hat{\Sigma}}_n^{\texttt{LZ}}\equiv\bm{\hat{\Sigma}}_n^{}(\mathbf{I}_{L_n})=\frac{1}{n}\sum_{g=1}^{G_n}\sum_{i,j\in\mathcal{T}_{g,n}}\mathbf{\hat{v}}_{i,n}\mathbf{\hat{v}}'_{j,n}\hat{u}_{i,n}\hat{u}_{j,n}.
\end{equation*}

The next theorem provides an asymptotic representation for this class of estimators.
\begin{theorem} Suppose Assumptions 1-3 hold.\\ If $\norm{\bm{\kappa}_{n}}_\infty=\max_{(g_1,i_1,j_1)}\sum_{g_2=1}^{G_n}\sum_{i_2,j_2\in\mathcal{T}_{g_2,n}}|\kappa_{g_1, g_2, i_1,j_1,i_2,j_2,n}|= O_p(1)$, then
\begin{equation}\label{expansion}\begin{split}
\bm{\hat{\Sigma}}_n^{\textup{\texttt{}}}(\bm{\kappa}_n)=\\ \frac{1}{n}\sum_{g_1=1}^{G_n}\sum_{g_2=1}^{G_n}\sum_{g_3=1}^{G_n}\sum_{i_1,j_1\in\mathcal{T}_{g_1,n}}\sum_{i_2,j_2\in\mathcal{T}_{g_2,n}}\sum_{i_3,j_3\in\mathcal{T}_{g_3,n}}\kappa_{g_1, g_2, i_1,j_1,i_2,j_2,n}&\mathbf{\hat{v}}_{i_1,n}\mathbf{\hat{v}}'_{j_1,n}M_{i_2j_3,n}M_{j_2i_3,n}\mathbb{E}[u_{i_3,n}u_{j_3,n}|\mathcal{X}_n,\mathcal{W}_n]\\
&+o_p(1).
\end{split}
\end{equation}
\end{theorem}
Heuristically, in Theorem 2 consistency of $\bm{\hat{\beta}}_n$ implies that the estimated residuals $\hat{u}_{i,n}$  asymptotically converge to $\tilde{u}_{i,n}=\sum_{j=1}^{n}M_{ij,n}u_{j,n}$, which are only affected by the estimation noise due to projecting out the high-dimensional covariates $\mathbf{w}_{i,n}$.

The result of Theorem 2 has a central importance in our analysis. First, it immediately provides an explicit characterization for the asymptotic limit of LZ's estimator, as shown in the following corollary.
\begin{corollary} Suppose the assumptions of Theorem 2 hold. Then,
\begin{equation*}
\bm{\hat{\Sigma}}_n^{\textup{\texttt{LZ}}}=\frac{1}{n}\sum_{g_1=1}^{G_n}\sum_{g_2=1}^{G_n}\sum_{i_1,j_1\in\mathcal{T}_{g_1,n}}\sum_{i_2,j_2\in\mathcal{T}_{g_2,n}}\mathbf{\hat{v}}_{i_1,n}\mathbf{\hat{v}}'_{j_1,n}M_{i_1j_2,n}M_{j_1i_2,n}\mathbb{E}[u_{i_2,n}u_{j_2,n}|\mathcal{X}_n,\mathcal{W}_n] + o_p(1).
\end{equation*}
\end{corollary}
Corollary 1 implies that inference based on LZ's clustered standard errors is invalid in general under asymptotics where $K_n/n\nrightarrow0$. In fact, $\bm{\hat{\Sigma}}_n^{\textup{\texttt{LZ}}}$ does not converge to the target $\bm{\Sigma}_n$ due to elements of $\mathbf{M}_n$ arising in its asymptotic limit. While the sign of the asymptotic bias of LZ's estimator cannot be determined in general, $\bm{\hat{\Sigma}}_n^{\textup{\texttt{LZ}}}$ will typically underestimate $\bm{\Sigma}_n$.\footnote{A particular case in which $\plim{\bm{\hat{\Sigma}}_n^{\textup{\texttt{LZ}}}}\leq \bm{\Sigma}_n$ holds in general is when the true residuals are in fact homoskedastic, which can be shown using arguments from Theorem 1 in \citet*{bellmccaffrey}.} Intuitively, the ``asymptotic" regression residuals $\tilde{u}_{i,n}$ tend to be smaller than the true residuals as a result of the overfitting due to the high-dimensional controls. In addition, estimated residuals will tend to have lower intra-cluster correlation than the true errors \citep*{bellmccaffrey}. Inference based on $\bm{\hat{\Sigma}}_n^{\textup{\texttt{LZ}}}$ is therefore expected to be asymptotically liberal in most applications.

Furthermore, Theorem 2 suggests that a particular choice of $\bm{\kappa}_n$ might set the leading term in the expansion (\ref{expansion}) equal to the target $\bm{\Sigma}_n$. Based on this insight, we define the estimator
\begin{equation*}
\bm{\hat{\Sigma}}_n^{\texttt{CR}}\equiv \bm{\hat{\Sigma}}^{\texttt{}}(\bm{\kappa}^{\texttt{CR}}_n)=\frac{1}{n}\sum_{g_1=1}^{G_n}\sum_{g_2=1}^{G_n}\sum_{i_1,j_1\in\mathcal{T}_{g_1,n}}\sum_{i_2,j_2\in\mathcal{T}_{g_2,n}}\kappa_{g_1, g_2, i_1,j_1,i_2,j_2,n}^{\texttt{CR}}\mathbf{\hat{v}}_{i_1,n}\mathbf{\hat{v}}'_{j_1,n}\hat{u}_{i_2,n}\hat{u}_{j_2,n},
\end{equation*}
 where $\bm{\kappa}^{\texttt{CR}}_n$ solves the system of $L_n(L_n-1)/2$  equations 
\begin{equation*}\begin{split}
\sum_{g_2=1}^{G_n}\sum_{i_2,j_2\in\mathcal{T}_{g_2,n}}\kappa_{g_1, g_2, i_1,j_1,i_2,j_2,n}M_{i_2,j_3,n}M_{j_2,i_3,n}&=\mathbbm{1}\{(g_1,i_1,j_1)=(g_3,i_3,j_3)\},\\ &1\leq g_1,g_3 \leq G_n,i_1,j_1 \in \mathcal{T}_{g_1,n},i_3,j_3 \in \mathcal{T}_{g_3,n}.
\end{split}
\end{equation*}
It also turns out that $\bm{\kappa}^{\texttt{CR}}_n$ can be characterized in closed form as
\begin{equation*}\label{kronecker}
\bm{\kappa}^{\texttt{CR}}_n=(\mathbf{S}_n'(\mathbf{M}_n\otimes\mathbf{M}_n) \mathbf{S}_n)^{-1},
\end{equation*}
where $\otimes$ denotes the Kronecker product and $\mathbf{S}_n$ is the $n^2 \times L_n$ selection matrix with full column rank such that $\mathbf{S}_n'\text{vec}(\bm{\Omega}_{u,n})$ is the $L_n\times 1$ vector containing the non-zero elements of $\bm{\Omega}_{u,n}$.
\begin{remark} When $\bm{\Omega}_{u,n}$ is assumed to be diagonal, i.e. errors are independent and (conditionally) heteroskedastic, then $G_n=n$, $\mathcal{T}_{i,n}=\{i\}$, $L_n=n$, $\mathbf{S}_n'(\mathbf{M}_n\otimes\mathbf{M}_n) \mathbf{S}_n=\mathbf{M}_n\odot\mathbf{M}_n$ where $\odot$ denotes the Hadamard product, and our estimator reduces to the heteroskedasticity-robust estimator of \citet{cjn}.
\end{remark}
In the next theorem we establish consistency of our proposed estimator.
\begin{theorem} 
Suppose Assumptions 1-3 hold.\\ If $\mathbb{P}[\lambda_{\text{min}}(\mathbf{S}_n'(\mathbf{M}_n\otimes\mathbf{M}_n) \mathbf{S}_n)>0]\to1$ and $\norm{\bm{\kappa}_{n}^{\textup{\texttt{CR}}}}_\infty= O_p(1)$, then
\begin{equation*}
\bm{\hat{\Sigma}}^{\textup{\texttt{CR}}}_n=\bm{\Sigma}_n +o_p(1).
\end{equation*}
\end{theorem}
Since $\mathbf{S}_n'(\mathbf{M}_n\otimes\mathbf{M}_n) \mathbf{S}_n$ is observable, the first high-level condition in Theorem 3 is expected to be verified whenever $\mathbf{S}_n'(\mathbf{M}_n\otimes\mathbf{M}_n) \mathbf{S}_n$ is invertible. The second high-level condition could be verified using Theorem 1 of \citet*{varah}, which provides a bound for $\norm{\bm{\kappa}_{n}^{\textup{\texttt{CR}}}}_\infty$ under the condition that $\mathbf{S}_n'(\mathbf{M}_n\otimes\mathbf{M}_n) \mathbf{S}_n$ is diagonally dominant. 
In simulations we find that diagonal dominance typically does not hold but our high-level condition is verified in a wide range of models and designs, as shown in Section 5.\footnote{\citet{cjn} instead develop their theory under the requirement that $\mathbf{M}_n\odot\mathbf{M}_n$ is diagonally dominant. It would be interesting to investigate whether this requirement could be relaxed in practice.}

\begin{remark}
Notice that explicit computation of $\bm{\kappa}^{\texttt{CR}}_n$ is not required for the purpose of variance estimation. Having defined $\mathbf{\hat{V}}_n=(\mathbf{\hat{v}}_{1,n},\dots,\mathbf{\hat{v}}_{n,n})'$ and $\mathbf{c}_n=(\mathbf{S}_n'(\mathbf{M}_n\otimes\mathbf{M}_n) \mathbf{S}_n)^{-1}\mathbf{S}_n(\mathbf{\hat{u}}_n\otimes\mathbf{\hat{u}}_n)$, one has $\vect(\bm{\hat{\Sigma}}^{\textup{\texttt{CR}}}_n)=(\mathbf{\hat{V}}_n\otimes\mathbf{\hat{V}}_n)'\mathbf{S}_n\mathbf{c}_n$. As a result, computing our variance estimator only requires to solve the linear system $(\mathbf{S}_n'(\mathbf{M}_n\otimes\mathbf{M}_n) \mathbf{S}_n)\mathbf{c}_n=\mathbf{S}_n(\mathbf{\hat{u}}_n\otimes\mathbf{\hat{u}}_n)$ for $\mathbf{c}_n$.\end{remark}

The structure of our proposed estimator is related to the cluster-robust variance estimator proposed by \citet*{bellmccaffrey}, which corresponds to a particular choice of block-diagonal $\bm{\kappa}_n$ that sets the the bias of the variance estimator to 0 only in the special case in which $\bm{\Omega}_{u,n}=\sigma^2\mathbf{I}_n$, i.e. the true residuals are in fact homoskedatic. Differently from \citet*{bellmccaffrey}, our choice of correction matrix $\bm{\kappa}^{\texttt{CR}}_n$ induces an averaging over cross-products of estimated residuals not just within but also across clusters, thus allowing to set the leading term in expansion (\ref{expansion}) equal to $\bm{\Sigma}_n$ in general.

The results of this paper can be easily extended to a more general version of the variance estimators, described in Section \ref{sectionF} of the Appendix, that allows to impose within-cluster zero restrictions on the variace-covariance matrix of the errors. In such form, our proposed estimator reduces to the one of \citet*{stockwatson2008} in the case of one-way fixed effects panel data models with zero restrictions on the conditional autocovariances of $U_{it}$ within entities. While our results cover a much wider class of models, they also partly improve on \citet*{stockwatson2008} as we do not require $(\mathbf{X}_{i1}',\dots, \mathbf{X}_{iT}', U_{i1}, \dots, U_{iT})$ to be i.i.d. nor we require $(\mathbf{X}_{it}, U_{it})$ to be stationary.

\subsection{Consistency of Liang and Zeger's estimator}
Although consistency of $\bm{\hat{\Sigma}}_n^{\textup{\texttt{CR}}}$ is derived under asymptotic sequences that allow but do not require $K_n/n\nrightarrow0$, it is still desirable to establish consistency of LZ's estimator under some sufficiently slow rate of growth for $K_n$. For this purpose, define $\mathbf{w}^{*}_{i,n}=\mathbf{w}_{i,n}\bm{\hat{\Sigma}}_{\mathbf{w},n}^{-1/2}$, where $\bm{\hat{\Sigma}}_{\mathbf{w},n}^{1/2}$ is the unique symmetric positive definite $K_n\times K_n$ matrix such that $\bm{\hat{\Sigma}}_{\mathbf{w},n}^{1/2}\bm{\hat{\Sigma}}_{\mathbf{w},n}^{1/2}=\frac{1}{n}\sum_{i=1}^{n}\mathbf{w}_{i,n}\mathbf{w}_{i,n}'$. The following theorem provides sufficient conditions for consistency of LZ's cluster-robust estimator.
\begin{theorem}
Suppose Assumptions 1-3 hold and that $\max_{i,j}\mathbb{E}[w_{ij,n}^{*2}]=O(1)$.
If $K_{n}^2/n\to0$, then
\begin{equation}\label{convergencewhit}
\bm{\hat{\Sigma}}^{\textup{\texttt{LZ}}}_n=\bm{\Sigma}_n +o_p(1).
\end{equation}
Moreover, if $\mathbb{E}[u^2_{i,n}|\mathcal{X}_n, \mathcal{W}_n]=\sigma_{n}^2$ $\forall i$, and $\mathbb{E}[u_{i,n}u_{j,n}|\mathcal{X}_n, \mathcal{W}_n]=0$ $\forall i\neq j$, then (\ref{convergencewhit}) holds under $K_n/n\to0$.
\end{theorem}
Although we can only prove consistency of LZ's estimator under $K_n^2/n\rightarrow0$, we speculate that $K_n/n\to0$ might suffice in general. We leave the refinement of this result for future work.\footnote{Theorem 4 also states that $K_n/n\to0$ is sufficient for consistency of LZ's estimator in the special case of homoskedastic errors. 
}

\section{Simulations}

This section reports the findings of a simulation study that investigates the finite sample behaviour of the cluster-robust variance estimators studied in this paper. We consider three distinct designs motivated by the empirical examples covered by the theoretical framework of this paper: the linear regression models with increasing dimension, the semiparametric partially linear model and the fixed effects panel data regression model.

\subsection{Results - Linear regression model with increasing dimension}
The chosen designs for our Monte Carlo experiments closely resemble those of \citet{cjn}, also borrowing from specifications in \citet*{stockwatson2008} and \citet*{mackinnon}. 
The data generating process (DGP) for the linear regression model with many covariates is:
\begin{equation}\label{modelmanycov}
\begin{split}
&y_{gi}=\beta x_{gi} + \bm{\gamma'}_n \mathbf{w}_{gi} + U_{gi}, \\&x_{gi}|\mathbf{w}_{gi}\sim \mathcal{N}(0,\sigma^2_{x,gi}), \quad \sigma_{x,gi}^2=\varkappa_{x}(1+(\bm{\iota}'\mathbf{w}_i)^2),\\
&U_{gi}=(\rho\mathbbm{1}(x_{gi}\geq 0)-\rho(1-\mathbbm{1}(x_{gi}\geq 0)) U_{g,i-1}+\varepsilon_{gi}, \quad \varepsilon_{gi}\sim\mathcal{N}(0,1), \\&u_{g1}\sim\mathcal{N}(0,\sigma_{u1}^{2}), \quad \sigma_{u1}^2=\varkappa_{u1}(1+(t(x_{g1})+\bm{\iota}'\mathbf{w}_{g1})^2),\\
& i=1,\dots,n/G, \quad g=1, \dots, G, \quad n=700,
\end{split}
\end{equation}
where $\mathbf{w}_{gi}\stackrel{i.i.d.}{\sim}\mathcal{U}(-1,1)$, $\bm{\iota}=(1,1, \dots,1)'$, $\beta=1$, $\bm{\gamma=0}$, $\rho=0.3 $, the constants $\varkappa_{x}$ and $\varkappa_{u1}$ are chosen so that $\mathbb{V}[x_{gi}]=\mathbb{V}[U_{g1}]=1$ and $t(a)=a\mathbbm{1}(-2\leq a \leq 2)+2\text{sgn}(a)(1-\mathbbm{1}(-2\leq a \leq 2))$.
\par Table \ref{manycov} reports the results of our experiment for five dimensions of $\mathbf{w}_{gi}$: $K \in \{1, 71, 141, 211, 281\}$, where the first covariate is an intercept, as well as three different numbers of equal-sized clusters: $G\in \{175, 70, 35 \}$. We consider three different estimators for the variance of the OLS estimator $\hat{\beta}$: the unfeasible estimator based on $\bm{\hat{\Sigma}}^{\texttt{Unf}}_n=\frac{1}{n}\sum_{g=1}^{G_n}\sum_{i,j\in\mathcal{T}_{g,n}}\mathbf{\hat{v}}_{i,n}\mathbf{\hat{v}}'_{j,n}U_{i,n}U_{j,n}$ that makes use of the true error realizations, the classical  estimator by LZ and our proposed cluster-robust formula, as previously defined. For each of these estimators, we report the bias (expressed in percentage), the standard deviation (denoted by Std.) and the empirical coverage probability (denoted by $\hat{p};\alpha$) of the Gaussian confidence interval of the form:
\begin{equation*}
\mathsf{l}_{\ell}\doteq\Bigg[\hat{\beta} - \Phi^{-1}(1-\alpha/2)\cdot\sqrt{\frac{\hat{\Omega}_{\ell}}{n}}, \hat{\beta} - \Phi^{-1}(\alpha/2)\cdot\sqrt{\frac{\hat{\Omega}_{\ell}}{n}} \Bigg], \quad \hat{\Omega}_{\ell}=\hat{\Gamma}^{-1}\hat{\Sigma}^{\ell}\hat{\Gamma}^{-1},
\end{equation*}
where $\Phi^{-1}$ denotes the inverse of the standard normal cumulative distribution function $\Phi$, $\hat{\Sigma}_{\ell}$ with $\ell\in\{\texttt{Unf, LZ, CR} \}$ corresponds to the variance estimators already discussed and we set $\alpha=0.05$.

\par The findings from this experiment are in line with our theoretical predictions. Firstly, we find that inference based on LZ's clustered standard errors formula is highly inaccurate. In fact, its bias quickly increases with the dimensionality of the model, resulting in substantial undercoverage even for $K/n=0.101$.
On the other hand, our proposed estimator performs well, with negligible bias and close-to-correct empirical coverage even for $K/n=0.401$. Such improvement in inference accuracy compared to LZ's estimator is achieved in spite of a decrease in relative precision. As expected, the performance of all estimators is adversely affected by a reduction in the number of clusters. In Tables \ref{absrowsum2}-\ref{absrowsum4} we also report on the behaviour of $\norm{\bm{\kappa}_{n}^{\texttt{CR}}}_\infty$ in this design; we find that $\norm{\bm{\kappa}^{\texttt{CR}}_{n}}_\infty$ does not only seem to be bounded but even decreasing as $n$ grows.\footnote{Notice that diagonal dominance of $\mathbf{S}_n'(\mathbf{M}_n\otimes\mathbf{M}_n) \mathbf{S}_n$ does not hold in any of the simulations carried out in this section.}

Analogous results are found for a different version of this experiment that considers independent and discrete controls constructed as $\mathbbm{1}\{\mathcal{N}(0,1)\geq 1\}$, as reported in Tables \ref{manycovdiscrete} and \ref{absrowsum2discrete}-\ref{absrowsum4discrete}.
\subsection{Results - Semiparametric partially linear model}\label{partiallysection}
\par
The experimental design chosen for the semiparametric partially linear model takes the form:
\begin{equation}\label{modelpartially}
\begin{split}
&y_{gi}=\beta x_{gi} + g(\mathbf{z}_{gi}) + U_{gi}, \\&x_{gi}=h(\mathbf{z}_{gi})+v_i, \quad v_{gi}|\mathbf{z}_{gi} \sim \mathcal{N}(0,\sigma^2_{v,gi}), \quad \sigma_{vi}^2=\varkappa_{v}(1+(\bm{\iota}'\mathbf{z}_{gi})^2),\\
&U_{gi}=(\rho\mathbbm{1}(z_{1,gi}\geq 0)-\rho(1-\mathbbm{1}(z_{1,gi}\geq 0)) U_{g,i-1}+\varepsilon_{gi}, \quad \varepsilon_{gi}\sim\mathcal{N}(0,1), \\&u_{g1}\sim\mathcal{N}(0,\sigma_{u1}^{2}), \quad \sigma_{u1}^2=\varkappa_{u1}(1+(t(x_{gi})+\bm{\iota}'\mathbf{z}_{gi})^2),\\
& i=1,\dots,n/G, \quad g=1, \dots, G, \quad n=700,
\end{split}
\end{equation}
where $\text{dim}(\mathbf{z}_{gi})=6$, $\mathbf{z}_{gi}=(z_{1,gi},\dots,z_{6,gi})'$ with $z_{\ell,gi}\stackrel{i.i.d.}{\sim}\mathcal{U}(-1,1)$, $\ell=1,\dots,6$. The unknown regressions functions are set to $g(\mathbf{z}_{gi})=\text{exp}\big(-\norm{\mathbf{z}_{gi}}^{1/2}\big)$ and $h(\mathbf{z}_{gi})=\text{exp}\big(\norm{\mathbf{z}_{gi}}^{1/2}\big)$, and the constants $\varkappa_{v}$ and $\varkappa_{u1}$ are again chosen so that $\mathbb{V}[x_{gi}]=\mathbb{V}[u_{g1}]=1$. Similarly to the previous simulation, we set  $\beta=1$ and $\rho=0.3$.  
\par To construct the covariates $\mathbf{w}_{gi}$ entering the estimated linear regression model $y_{gi}=\bm{\beta}'\mathbf{x}_{gi} + \bm{\gamma}_n'\mathbf{w}_{gi} + u_{gi}$, we consider power series expansions. The table below gives a summary of the expansions considered, where $\mathbf{w}_{gi}=\mathbf{p}(\mathbf{z}_{gi};K)$ for $K \in \{1, 7, 13, 28, 34, 84, 90, 210, 216\}$ is defined as follows:

\renewcommand{\arraystretch}{1}
\begin{table}[H]

\centering
\caption{Polynomial Basis Expansion: $\text{dim}(\mathbf{z}_{gi})=6$ and $n=700$}
\label{expansions}

\addtolength{\tabcolsep}{+5pt} 
\renewcommand{\arraystretch}{1.5} 
\begin{threeparttable}
\begin{tabular}{ccc}
\hline
$K$ & $\mathbf{p}(\mathbf{z}_{gi};K)$ & $K/n$\\ \toprule
1	& 1	& 0.001 \\
7	& $(1, z_{1,gi}, z_{2,gi}, z_{3,gi}, z_{4,gi}, z_{5,gi}, z_{6,gi})'$	& 0.010   \\
13	& $(\mathbf{p}(\mathbf{z}_{gi};7)', z_{1,gi}^2, z_{2,gi}^2, z_{3,gi}^2, z_{4,gi}^2, z_{5,gi}^2, z_{6,gi}^2)'$ & 0.019   \\
28	& $\mathbf{p}(\mathbf{z}_{gi};13)$ + first-order interactions & 0.040   \\
34	& $(\mathbf{p}(\mathbf{z}_{gi};28)', z_{1,gi}^3, z_{2,gi}^3, z_{3,gi}^3, z_{4,gi}^3, z_{5,gi}^3, z_{6,gi}^3)'$ & 0.049   \\
84	& $\mathbf{p}(\mathbf{z}_{gi};13)$ + second-order interactions & 0.120   \\
90	& $(\mathbf{p}(\mathbf{z}_{gi};84)', z_{1,gi}^4, z_{2,gi}^4, z_{3,gi}^4, z_{4,gi}^4, z_{5,gi}^4, z_{6,gi}^4)'$  & 0.129   \\
210	& $\mathbf{p}(\mathbf{z}_{gi};90)$ + third-order interactions & 0.300   \\
216	& $(\mathbf{p}(\mathbf{z}_{gi};210)', z_{1,gi}^5, z_{2,gi}^5, z_{3,gi}^5, z_{4,gi}^5, z_{5,gi}^5, z_{6,gi}^5)'$ & 0.309   \\
\bottomrule
\end{tabular}
\begin{tablenotes}
\footnotesize 
\item Source: \citeauthor*{cjn} (\citeyear{cjn}, Supplemental Appendix).
\end{tablenotes}
\end{threeparttable}
\end{table}

The results for this experiment are given in Table \ref{partiallymonte}, in which we only report $K\in \{1, 13, 34, 90, 216\}$ for reasons of parsimony. The numerical findings are largely consistent with those reported for the other two simulation models. Although $\norm{\bm{\kappa}_{n}^{\texttt{CR}}}_\infty$ has bigger magnitude in this setting compared to the other simulation models, it still appears to be bounded (see Tables \ref{absrowsumsemi2}-\ref{absrowsumsemi4}). 

The main difference between this setting and the linear model with increasing dimension considered previously is that the unfeasible estimator that uses realizations of the true structural disturbances is free not just from estimation error but also specification error, which in turn affects LZ's and our proposed estimator when $K$ is small; in addition, the degree of heteroskedasticity and dependence in the errors is invariant with respect to the dimensionality of the model, since it only depends on $x_{gi}$ and $\mathbf{z}_{gi}$ but not $\mathbf{w}_{gi}$.

\subsection{Results - Fixed effects panel data regression model}

For fixed effects panel data regression model we consider the following specification:

\begin{equation}\label{twowayoriginal}
y_{it}=\beta x_{it} +\alpha_ i + e_{d_{it}}+ u_{it}, \qquad i=1,\dots, N, \qquad t=1,\dots, T,
\end{equation}
where $\alpha_i$ is a time-invariant individual effect and $e_{d_{it}}$ are unobserved factors common to all observations sharing the same value of the indexing variable $d_{it}\in\{1,\dots, N_d\}$. This model coincides with the one studied in Verdier (2018), whose theory and simulation results concern the case of two-way clustering. We instead consider the case of one-way clustering at the individual level as we postulate the following DGP:
\begin{equation}\label{twowaydgp}
\begin{split}
&y_{it}=\beta x_{it} +\alpha_ i + e_{d_{it}} + U_{it}, \\&x_{it}|\mathbf{z}_{it}\sim \mathcal{N}(0,\sigma^2_{x,it}), \quad \sigma_{x,gi}^2=\varkappa_{x}(1+(\bm{\iota}'\mathbf{z}_{it})^2),\\
&U_{it}=(\rho\mathbbm{1}(x_{it}\geq 0)-\rho(1-\mathbbm{1}(x_{it}\geq 0)) U_{i,t-1}+\varepsilon_{it}, \quad \varepsilon_{it}\sim\mathcal{N}(0,1), \\&u_{i1}\sim\mathcal{N}(0,\sigma_{u1}^{2}), \quad \sigma_{u1}^2=\varkappa_{u1}(1+(t(x_{i1})+\bm{\iota}'\mathbf{z}_{i1})^2),\\
& i=1,\dots,N, \quad t=1, \dots, T,
\end{split}
\end{equation}
 where $\text{dim}(\mathbf{z}_{it})=6$, $\mathbf{z}_{it}=(z_{1,it},\dots,z_{6,it})'$ with $z_{\ell,gi}\stackrel{i.i.d.}{\sim}\text{Uniform}(-1,1)$, $\ell=1,\dots,6$,  the constants $\varkappa_{x}$ and $\varkappa_{u1}$ are chosen so that $\mathbb{V}[x_{it}]=\mathbb{V}[U_{i1}]=1$, the function $t(\cdot)$ is as previously defined and we set $\beta=1$ and $\alpha_i=e_{d_{it}}=0$. For the purpose of estimation, we transform (\ref{twowayoriginal}) by partialling out the individual fixed effects $\alpha_i$, so that the estimated model $\tilde{y}_{it}=\beta \tilde{x}_{it} + \tilde{e}_{d_{it}}+ \tilde{u}_{it}$ has $\text{dim}(\mathbf{w}_{i})=N_d$.\footnote{The motivation for this transformation is that $(\mathbf{S}_n'(\mathbf{M}_n\otimes\mathbf{M}_n) \mathbf{S}_n)$ is not invertible when the controls $\mathbf{w}_{i,n}$ include indicators for the clusters (see, e.g., \citealp*{stockwatson2008}). Notice that partialling out the fixed effects does not affect the correlation structure of the errors.} We consider $G=N=[700/T]$ for $T\in\{4, 10, 20\}$, as well as $N_d=700/r$ for $r\in\{700, 10, 5, 4, 3\}$, so that the total sample size is always roughly $n=700$. Tables \ref{manycovtwoway} and \ref{absrowsumtwoway2}-\ref{absrowsumtwoway3} report the numerical findings of this experiment, which are consistent with our theoretical predictions and in line with the results obtained for the other simulation models.

\section{Empirical Illustration}

In this section we illustrate the use of the inference methods discussed in this paper by revisiting \citet*{dl2001} study of the impact of abortion on crime rates.  

\citeauthor*{dl2001} (\citeyear{dl2001}, henceforth DL) put forward the hypothesis that the legalization of abortion in the United States in the 1970s played a major role in explaining the sharp decline in crime observed two decades later. In particular, they describe two causal channels through which abortion might affect crime. The first is that abortion reduces the absolute size of a cohort, resulting in lower crime 15-25 years later, when its members are at the highest risk of engaging in criminal activities. The second channel is ascribed to the increased control over fertility that abortion provides to women. In fact, women may use abortion to optimize the timing of childbearing, thus ensuring that the child grows in a more favourable environment, e.g. when a father is present in the family, the mother is better educated and household income is stable. As a result, increased access to abortion is expected to cause a reduction in crime levels even if fertility rates were to remain constant.

In order to estimate the impact of abortion on crime, \citet{dl2001} consider state-level yearly data for the period 1985-1997 and propose a model for crime rates whose basic specification is 

\begin{equation}{\label{DL}}
y_{cit}=\beta_{c}a_{cit}+\bm{\delta}_{c}'\mathbf{z}_{it} + \theta_{ci} + \lambda_{ct} + u_{cit},
\end{equation}
where $i$ indexes states, $t$ indexes the time period, $c\in\{\text{violent, property, murder}\}$ indexes the type of crime, $y_{cit}$ is the crime-rate for crime type $c$; $a_{cit}$ is measure of abortion rate relevent for crime type $c$; $z_{it}$ is a set of time-varying state-specific controls consisting of the log of lagged prisoners per capita, the log of lagged police per capita, the unemployment rate, per-capita income, the poverty rate, AFDC generosity at time $t-15$, a dummy for concealed weapons law and beer consumptions; $\theta_{ci}$ are state fixed effects; and $\lambda_{ct}$ are time fixed effects. Further details on data definitions and the institutional background can be found in the original paper.

The results from estimating the baseline model in (\ref{DL}) are reported in Table \ref{empirical findings} and resemble those in \citet{dl2001}, although not identical as we have excluded Washington DC from the sample.\footnote{We exclude Washington DC for simplicity, as it produces similar results to \citet{dl2001} and circumvents the need to introduce the estimation weights used in their paper.} Following \citet{dl2001}, we report standard errors clustered at the state level.
These estimates indicate a strong (and statistically significant) negative association between abortion and crime, as they imply that an increase in the abortion rate of 100 per 1,000 live births is associated with a reduction in crime rates between 9 and 13 per cent, depending on the type of crime. However, the extent to which this association can be interpreted as causal crucially depends on the assumption that abortion rates can be taken as random after controlling for a national trend, time-invariant state-specific confounders and $z_{it}$. 
 Even if one believes that abortion rates can be taken as exogenous conditional on the the controls included by \citet{dl2001}, one can still expect the assumption that they enter the structural equation for crime rates linearly as in (\ref{DL}) to be too restrictive. For example, \citet*{foote} have argued that the results in \citet{dl2001} might not be robust to the inclusion of state-specific trends.\footnote{In a response to \citet*{foote}, \citet*{dl2008} reexamine their orginal study and use a longer panel to argue that their original results are robust to the inclusion of state-specific linear time trends.}

For these reasons, we consider a model for crime rates and abortion in which the controls $\mathbf{z}_{it}$ are allowed to enter in a much more flexible way compared to \citet{dl2001}. In particular, we consider a version of the high-dimensional regression model studied in this paper where in addition to the controls included by \citet{dl2001}, we include first-order interactions, quadratics, cumulative values and interactions of those variables and their initial values with a quadratic trend; in addition, we also include the interaction between the initial level of abortion and a quadratic trend. Once we stack all these regressors and the time-effects $\lambda_{ct}$ in the vector $\mathbf{w}_{it}$ and absorb the state-effects, we obtain a regression model of the same form as (\ref{model}):
\begin{equation}\label{DLhighdimensional}
\tilde{y}_{cit}=\beta_{c}\tilde{a}_{cit}+\bm{\gamma}_{c}'\mathbf{\tilde{w}}_{it} + \tilde{u}_{cit},
\end{equation}
where the dimension of the high-dimensional controls is $K_n=105$, resulting in $K_n/n\approx0.161$. 
Estimates for the causal effect of abortion on crime based on (\ref{DLhighdimensional}) are given in Table \ref{empirical findings}, where we also report standard errors based on the variance estimators considered in this paper. These estimates are qualitatively similar to those obtained for the baseline model considered in \citet{dl2001}, and interestingly imply an even more sizeable negative effect of abortion on crime rates for all types of crime. The statistical significance of these effects however crucially depends on the choice of standard errors. In fact, clustered standard errors based on the variance estimator proposed in this paper are between 42 and 74 per cent bigger than the traditional clustered stardard errors by \citet{liangzeger}, depending on the type of crime. In the case of violent crime, for example, the estimated coefficient for abortion rates has associated p-value below 1 per cent when traditional clustered standard errors are used, while the use of our proposed standard errors leads to failure to reject the hypothesis of no effect of abortion on crime at the 5 per cent level.

This empirical illustration showcases the relevance of the inference methods proposed in this paper. In this particular application, the inclusion of many controls arises naturally as a way to flexibly control for observable state-level characteristics and trends that are allowed to depend on those characteristics.
Our approach in this particular application resembles the one adopted by \citet{belloni2014rev}, who also re-examine the empirical setting in \citet{dl2001} to illustrate the use of their proposed inference method for treatment effects with many controls based on LASSO double-selection. They consider a similar specification of the high-dimensional model in (\ref{DLhighdimensional}) but allow for an even more flexible specification that includes higher-order interactions of the variables we consider (and a few additional ones, such as initial differences of $\mathbf{z}_{it}$) with cubic trends, which gives $K_n/n\approx0.500$ in their application.\footnote{Interestingly, their estimates imply statistically non-significant impact of abortion on all types of crime.} While  \citeauthor{belloni2014rev}'s (\citeyear{belloni2014rev}) method is naturally suited to handle such large number of controls, its validity relies on the assumption that the effect of confounding factors can be controlled for by a small number of variables (``approximate sparsity"). 
Our proposed inference procedure therefore offers a valuable alternative to selection-based methods in settings where the inclusion of a relatively large number of controls is expected to yield a reasonable approximation of the structural relationship of interest, while circumventing the need to impose requirements of sparsity on the model.

\section{Conclusion}

This paper provides inference results for the OLS estimator of a subset of coefficients in linear regression models with many controls and clustering. We show that the usual cluster-robust variance estimator by \citet*{liangzeger} does not deliver consistent standard errors when the number of controls is a non-vanishing fraction of the sample size, typically resulting in confidence intervals with coverage below the nominal size. We then propose a new clustered standard error formula that is robust to the inclusion of many controls. Monte Carlo evidence supports our theoretical results and shows that our proposed variance estimator performs well in finite samples.

While our results are presented for the case of one-way clustering, we expect that they can be easily adapted to the generalisation of our methods to multi-way clustering proposed by \citet*{verdier}. It would also be of interest to investigate whether the analysis of this paper could be extended to cases where variance estimation does not rely on zero restrictions on the covariance matrix of the errors, e.g. when time series or spacial dependence in the errors is assumed.

\newpage

\begin{landscape}

\renewcommand{\arraystretch}{1.3}
\begin{table}[h]
\centering
\caption{Monte Carlo simulations for linear regression model with increasing dimension (continuous controls), $n=700$}
\label{manycov}

\addtolength{\tabcolsep}{-0.8pt} 
\begin{threeparttable}
\begin{tabular}{lccccccccccc}
\hline
& \multicolumn{2}{c}{$\hat{\beta}$} & \multicolumn{3}{c}{Unfeasible} & \multicolumn{3}{c}{Classical} &\multicolumn{3}{c}{Robust}\\ \cmidrule(l{1pt}r{1pt}){2-3} \cmidrule(l{1pt}r{1pt}){4-6} \cmidrule(l{1pt}r{1pt}){7-9} \cmidrule(l{1pt}r{1pt}){10-12}
&Mean&Variance&Bias (\%)&Std.&$\hat{p};.05$ &Bias (\%)&Std.&$\hat{p};.05$&Bias (\%)&Std.&$\hat{p};.05$\\ \toprule
\multicolumn{2}{@{}c}{\makecell{\boldmath$G=175$}}\\
$\quad K/n=0.001$ & 1.00 & .0042 & 1.29 & .0011  & .047 & -.101 & .0011&.051&.360&.0011 &.051\\
$\quad K/n=0.101$    &  1.00   & .752 & -1.42& .492 &.044& -19.3&.375&.080&-4.45&.460&.055\\
$\quad K/n=0.201$   & .981  &2.89  & -.578 & 2.09 & .043&-19.6&1.21&.106&-4.41&1.87&.053 \\
$\quad K/n=0.301$   & 1.00  &6.56  & 2.04 & 4.18 & .042& -38.0&2.02&.143&-1.71&3.93&.054 \\
$\quad K/n=0.401$   & 1.02  &11.7  & -1.42 & 7.28 & .044&-53.2&2.76&.179&-6.17&6.76&.057  \\ \hline
\multicolumn{2}{@{}c}{\makecell{\boldmath$G=70$}}\\
$\quad K/n=0.001$   &1.00 	& .0027& -2.00	&$7.93{\times}10^{-4}$ &.053&-4.02&$7.73{\times}10^{-4}$&.064&-3.84&$7.75{\times}10^{-4}$&.064\\
$\quad K/n=0.101$   &1.01  	&.310  & -3.16 	& .317 & .039&-21.4&.235&.072&-7.15&.289&.052\\
$\quad K/n=0.201$   &1.02  	&1.10  & 2.45 	& 1.15 & .036&-30.0&.702&.096&-1.80&1.07&.050 \\
$\quad K/n=0.301$   &1.04  	&2.60  & 1.77 	&2.67&.034& -42.5&1.273&.133& -3.32 &2.48&.052 \\
$\quad K/n=0.401$   &1.01  	&4.82  & -4.27 	& 4.50&.039 & -54.7& 1.690&.181 &-10.3&4.12&.058  \\ \hline
\multicolumn{2}{@{}c}{\makecell{\boldmath$G=35$}}\\
$\quad K/n=0.001$   & 1.00 & .0021 & .51& $7.00{\times}10^{-4}$&.047& -2.86&$6.80{\times}10^{-4}$ &.062&-2.70&$6.82{\times}10^{-4}$&.062\\
$\quad K/n=0.101$   &1.00  &.156  & 1.71 & .027 & .037&-18.3&.194&.065&-4.3& .238&.045\\
$\quad K/n=0.201$   &1.01  &.558 & 1.73  & .858 & .027 & -31.6&.510&.089&-5.12&.770&.048 \\
$\quad K/n=0.301$   & 1.00 &1.33  & -.16 &1.81 &.033& -45.6&.840&.140 &-9.10& 1.61 &.070 \\
$\quad K/n=0.401$   & .973  &2.41  & .70 & 3.05 & .037& -53.7&1.12&.186&-9.50&2.72&.083  \\ \hline
\bottomrule
\end{tabular}
\begin{tablenotes}
\footnotesize 
\item Notes: Simulation results based on 5,000 replications. DGP as described in Equation \ref{modelmanycov}.
\end{tablenotes}
\end{threeparttable}
\end{table}
\end{landscape}

\begin{landscape}

\renewcommand{\arraystretch}{1.3}
\begin{table}[h]
\centering
\caption{Monte Carlo simulations for linear regression model with increasing dimension (discrete controls), $n=700$}
\label{manycovdiscrete}

\addtolength{\tabcolsep}{-0.8pt} 
\begin{threeparttable}
\begin{tabular}{lccccccccccc}
\hline
& \multicolumn{2}{c}{$\hat{\beta}$} & \multicolumn{3}{c}{Unfeasible} & \multicolumn{3}{c}{Classical} &\multicolumn{3}{c}{Robust}\\ \cmidrule(l{1pt}r{1pt}){2-3} \cmidrule(l{1pt}r{1pt}){4-6} \cmidrule(l{1pt}r{1pt}){7-9} \cmidrule(l{1pt}r{1pt}){10-12}
&Mean&Variance&Bias (\%)&Std.&$\hat{p};.05$ &Bias (\%)&Std.&$\hat{p};.05$&Bias (\%)&Std.&$\hat{p};.05$\\ \toprule
\multicolumn{2}{@{}c}{\makecell{\boldmath$G=175$}}\\
$\quad K/n=0.001$ & 1.00 & .0043 & -.46 & .0011  & .048 &-1.70 & .0011&.051&-1.50&.0011 &.051\\
$\quad K/n=0.101$    &  1.00   & .0019 & -1.84& $3.01{\times}10^{-4}$ &.052& -18.4&$2.60{\times}10^{-4}$&.072&-3.03&$3.21{\times}10^{-4}$&.057\\
$\quad K/n=0.201$   & 1.00  &.0020  & 1.63 & $3.10{\times}10^{-4}$ & .048&-20.9&$2.39{\times}10^{-4}$&.086&-.410&$3.60{\times}10^{-4}$&.053 \\
$\quad K/n=0.301$   & 1.00  &.0023  & -3.40 & $3.52{\times}10^{-4}$ & .055& -34.2&$2.41{\times}10^{-4}$&.114&-5.20&$4.53{\times}10^{-4}$&.064 \\
$\quad K/n=0.401$   & 1.00  &.0026  & -1.00 & $4.05{\times}10^{-4}$ & .050&-41.5&$2.44{\times}10^{-4}$&.140&-2.48&$5.88{\times}10^{-4}$&.061  \\ \hline
\multicolumn{2}{@{}c}{\makecell{\boldmath$G=70$}}\\
$\quad K/n=0.001$   &1.00 	& .0026& -.161	&$8.07{\times}10^{-4}$ &.046&-2.17&$7.88{\times}10^{-4}$&.056&-1.99&$7.91{\times}10^{-4}$&.056\\
$\quad K/n=0.101$    &  1.00   & .0018 & -1.48& $3.72{\times}10^{-4}$ &.049& -14.4&$3.27{\times}10^{-4}$&.075&-3.88&$4.07{\times}10^{-4}$&.059\\
$\quad K/n=0.201$   & 1.00  &.0020  & -.773 & $4.00{\times}10^{-4}$ & .052&-22.9&$3.16{\times}10^{-4}$&.091&-3.26&$4.87{\times}10^{-4}$&.062 \\
$\quad K/n=0.301$   & 1.00  &.0024  & -4.56 & $4.60{\times}10^{-4}$ & .054& -35.3&$3.13{\times}10^{-4}$&.124&-8.16&$6.11{\times}10^{-4}$&.071 \\
$\quad K/n=0.401$   & 1.00  &.0027  & -1.16 & $5.20{\times}10^{-4}$ & .053&-42.5&$3.13{\times}10^{-4}$&.146&-5.96&$8.00{\times}10^{-4}$&.076  \\ \hline
\multicolumn{2}{@{}c}{\makecell{\boldmath$G=35$}}\\
$\quad K/n=0.001$   & 1.00 & .0021 & .233& $7.26{\times}10^{-4}$&.046& -2.97&$7.06{\times}10^{-4}$ &.063&-2.97&$7.07{\times}10^{-4}$&.063\\
$\quad K/n=0.101$    &  1.00   & .0018 & 1.29& $4.85{\times}10^{-4}$ &.046& -13.0&$4.19{\times}10^{-4}$&.075&-2.77&$5.21{\times}10^{-4}$&.062\\
$\quad K/n=0.201$   & 1.00  &.0020  & .570 & $5.32{\times}10^{-4}$ & .046&-23.1&$4.11{\times}10^{-4}$&.099&-4.24&$6.37{\times}10^{-4}$&.069 \\
$\quad K/n=0.301$   & 1.00  &.0023  & -.892 & $6.08{\times}10^{-4}$ & .052& -34.2&$4.07{\times}10^{-4}$&.122&-8.16&$8.03{\times}10^{-4}$&.080 \\
$\quad K/n=0.401$   & 1.00  &.0028  & -4.11 & $6.97{\times}10^{-4}$ & .055&-45.4&$4.06{\times}10^{-4}$&.160&-13.9&$11.0{\times}10^{-4}$&.102  \\ \hline
\bottomrule
\end{tabular}
\begin{tablenotes}
\footnotesize 
\item Notes: Simulation results based on 5,000 replications. DGP as described in Equation \ref{modelmanycov}, with $w_{\ell,gi}=\mathbbm{1}\{\mathcal{N}(0,1,)\geq 1\},\quad \forall \ell, g, i$.
\end{tablenotes}
\end{threeparttable}
\end{table}
\end{landscape}

\begin{landscape}

\renewcommand{\arraystretch}{1.3}
\begin{table}[h]
\centering
\caption{Monte Carlo simulations for semiparametric partially linear model, $n=700$}
\label{partiallymonte}

\addtolength{\tabcolsep}{+1pt} 
\begin{threeparttable}
\begin{tabular}{lccccccccccc}
\hline
& \multicolumn{2}{c}{$\hat{\beta}$} & \multicolumn{3}{c}{Unfeasible} & \multicolumn{3}{c}{Classical} &\multicolumn{3}{c}{Robust}\\ \cmidrule(l{1pt}r{1pt}){2-3} \cmidrule(l{1pt}r{1pt}){4-6} \cmidrule(l{1pt}r{1pt}){7-9} \cmidrule(l{1pt}r{1pt}){10-12}
&Mean&Variance&Bias (\%)&Std.&$\hat{p};.05$ &Bias (\%)&Std.&$\hat{p};.05$&Bias (\%)&Std.&$\hat{p};.05$\\ \toprule
\multicolumn{2}{@{}c}{\makecell{\boldmath$G=175$}}\\
$\quad K/n=0.001$ 	& .984 	& .0436	&-1.42	&.0192	&.048	&-2.84	&.0182	&.053	&-2.70	&.0183	&.052\\
$\quad K/n=0.019$   &1.00	&.0519 	&-1.80	&.0231	&.046	&-5.33	&.0210	&.052	&-3.23	&.0220	&.049\\
$\quad K/n=0.049$   & .994  &.0515  & 2.22 	&.0225	&.044	&-5.14	&.0195	&.056	&.680	&.0219	&.049 \\
$\quad K/n=0.129$   & .993  &.0570	& .922	&.0231	&.050	&-15.0	&.0174	&.078	&-.760	&.0230	&.056 \\
$\quad K/n=0.309$   & .993  &.0733  & .824 	&.0287	&.049	&-31.0	&.0170	&.103	&-1.94	&.0304	&.056  \\ \hline
\multicolumn{2}{@{}c}{\makecell{\boldmath$G=70$}}\\
$\quad K/n=0.001$ 	& .986 	& .0433	&-.89	&.0196	&.050	 &-2.31	 &.0189	 &.054	&-2.20	&.0183&.054\\
$\quad K/n=0.019$   &1.00	&.0209 	&.28	&.0132	&.042	&-4.13	&.0120	&.052&-2.08	&.0125	&.050\\
$\quad K/n=0.049$   & 1.00  &.0218  & -.59 	&.0136 	&.041	&-8.36	&.0117	&.0548	&-3.07	&.0130	&.049 \\
$\quad K/n=0.129$   & 1.00  &.0241  & -.04	&.0153	&.046	&-16.5	&.0113	&.074	&-2.87	&.0148	&.059 \\
$\quad K/n=0.309$   & 1.00  &.0315  & -1.44 &.0172 	&.043	&-33.0	&.0100	&.115	&-5.68	 &.0174	&.067  \\ \hline
\multicolumn{2}{@{}c}{\makecell{\boldmath$G=35$}}\\
$\quad K/n=0.001$ 	& .988 	& .0098	&-1.37	&.0083	&.042	 &-5.01	 &.0079	 &.059	&-4.88	&.0079	&.059\\
$\quad K/n=0.019$   &1.00	&.0121 	&-3.54	&.0108	&.050	&-9.40	&.0095	&.066&-3.41	&.0099&.061\\
$\quad K/n=0.049$   & 1.00  &.0116  & -.41 & .0092 & .038&-9.70&.0079&.064&-4.64&.0087&.060 \\
$\quad K/n=0.129$   & 1.00  &.0127  & 1.70 & .0100 & .033& -16.0&.0073&.070&-3.02&.0096&.058 \\
$\quad K/n=0.309$   & 1.00  &.0169  & 1.29 & .0134 & .040&-31.9	&.0074&.115&-6.50&.0118&.065  \\ \hline
\bottomrule
\end{tabular}
\begin{tablenotes}
\footnotesize 
\item Notes:  Simulation results based on 5,000 replications. DGP as described in Equation \ref{modelpartially}.
\end{tablenotes}
\end{threeparttable}
\end{table}
\end{landscape}

\begin{landscape}

\renewcommand{\arraystretch}{1.3}
\begin{table}[h]
\centering
\caption{Monte Carlo simulations for two-way fixed effects panel data regression model, $n=700$}
\label{manycovtwoway}

\addtolength{\tabcolsep}{-0.8pt} 
\begin{threeparttable}
\begin{tabular}{lccccccccccc}
\hline
& \multicolumn{2}{c}{$\hat{\beta}$} & \multicolumn{3}{c}{Unfeasible} & \multicolumn{3}{c}{Classical} &\multicolumn{3}{c}{Robust}\\ \cmidrule(l{1pt}r{1pt}){2-3} \cmidrule(l{1pt}r{1pt}){4-6} \cmidrule(l{1pt}r{1pt}){7-9} \cmidrule(l{1pt}r{1pt}){10-12}
&Mean&Variance&Bias (\%)&Std.&$\hat{p};.05$ &Bias (\%)&Std.&$\hat{p};.05$&Bias (\%)&Std.&$\hat{p};.05$\\ \toprule
\multicolumn{2}{@{}c}{\makecell{\boldmath$G=175$}}\\
$\quad K/n=0.001$ & 1.00 & .0025 & .435 & $7.02{\times}10^{-4}$  & .045 &-1.28 & $6.72{\times}10^{-4}$&.053&-1.28&$6.72{\times}10^{-4}$ &.053\\
$\quad K/n=0.100$    &  1.00   & .0029 & -1.26& $7.74{\times}10^{-4}$ &.048& -17.6&$6.07{\times}10^{-4}$&.076&-3.08&$7.82{\times}10^{-4}$&.056\\
$\quad K/n=0.200$    & 1.00  &.0033  & .350 & $8.30{\times}10^{-4}$ & .046& -30.4&$5.18{\times}10^{-4}$&.105&-2.02&$8.80{\times}10^{-4}$&.056 \\
$\quad K/n=0.250$    & 1.00  &.0036  & -1.22 & $8.71{\times}10^{-4}$ & .052& -38.0&$5.04{\times}10^{-4}$&.127&-3.86&$9.95{\times}10^{-4}$&.062 \\
$\quad K/n=0.333$    & 1.00  &.0042  & -1.51 & $9.61{\times}10^{-4}$ & .051&-48.8&$4.63{\times}10^{-4}$&.167&-5.18&$.0012$&.066  \\ \hline
\multicolumn{2}{@{}c}{\makecell{\boldmath$G=70$}}\\
$\quad K/n=0.001$   &1.00 	& .0020& -2..30&$5.15{\times}10^{-4}$ &.049&-4.60&$5.04{\times}10^{-4}$&.059&-4.50&$5.05{\times}10^{-4}$&.059\\
$\quad K/n=0.100$   &  1.00   & .0022 & -1.58& $5.83{\times}10^{-4}$ &.049& -15.4&$4.89{\times}10^{-4}$&.075&-3.99&$6.04{\times}10^{-4}$&.061\\
$\quad K/n=0.200$   & 1.00  &.0024  & -.482 & $6.20{\times}10^{-4}$ & .046&-25.9&$4.41{\times}10^{-4}$&.100&-4.13&$6.92{\times}10^{-4}$&.062 \\
$\quad K/n=0.250$   & 1.00  &.0026  & .498 & $6.53{\times}10^{-4}$ & .045& -30.8&$4.32{\times}10^{-4}$&.111&-3.63&$7.66{\times}10^{-4}$&.063 \\
$\quad K/n=0.333$   & 1.00  &.0029  & -1.84 & $7.13{\times}10^{-4}$ & .051&-41.2&$4.11{\times}10^{-4}$&.136&-6.55&$9.14{\times}10^{-4}$&.071  \\ \hline
\multicolumn{2}{@{}c}{\makecell{\boldmath$G=35$}}\\
$\quad K/n=0.001$   & 1.00 & .0018 & -3.66& $5.26{\times}10^{-4}$&.049& -7.32&$5.07{\times}10^{-4}$ &.068&-7.19&$5.08{\times}10^{-4}$&.068\\
$\quad K/n=0.100$    &  1.00   & .0020 & -1.98& $5.87{\times}10^{-4}$ &.046& -16.1&$4.98{\times}10^{-4}$&.080&-6.30&$4.98{\times}10^{-4}$&.066\\
$\quad K/n=0.200$   & 1.00  &.0022  & -1.21 & $6.22{\times}10^{-4}$ & .044&-26.0&$4.65{\times}10^{-4}$&.099&-6.70&$7.20{\times}10^{-4}$&.068 \\
$\quad K/n=0.250$   & 1.00  &.0023  & -.057 & $6.83{\times}10^{-4}$ & .045& -29.7&$4.83{\times}10^{-4}$&.109&-5.45&$8.41{\times}10^{-4}$&.073 \\
$\quad K/n=0.333$   & 1.00  &.0027  & -2.90 & $7.21{\times}10^{-4}$ & .050&-40.8&$4.43{\times}10^{-4}$&.145&-10.7&$9.72{\times}10^{-4}$&.093  \\ \hline
\bottomrule
\end{tabular}
\begin{tablenotes}
\footnotesize 
\item Notes: Simulation results based on 5,000 replications. DGP as described in Equation \ref{twowaydgp}.
\end{tablenotes}
\end{threeparttable}
\end{table}
\end{landscape}

\begin{table}[h!]

\centering
\caption{Absolute row sum of $\bm{\kappa}_n$ for $K_n/n=0.200$ - Linear regression model with many continuous controls}
\label{absrowsum2}
\addtolength{\tabcolsep}{+5pt} 
\begin{threeparttable}
\begin{tabular}{lccc}
\hline
\multicolumn{3}{@{}c}{$\bm{G=140}$}\\

$n=250$ & $n=500$ & $n=750$ & $n=1000$  \\
\midrule
\shortstack{$4.93$ \\ $(0.26)$}& \shortstack{$4.23$ \\ $(0.14)$}   & \shortstack{$3.95$ \\ $(0.11)$}  & \shortstack{$3.80$ \\ $(0.090)$} \\ 
\midrule
\multicolumn{3}{@{}c}{$\bm{G=70}$}\\

$n=250$ & $n=500$ & $n=750$ & $n=1000$  \\
\midrule
\shortstack{$8.41$ \\ $(0.56)$}& \shortstack{$6.71$ \\ $(0.24)$}   & \shortstack{$6.11$ \\ $(0.18)$}  & \shortstack{$5.81$ \\ $(0.14)$} \\ 
\midrule
\multicolumn{3}{@{}c}{$\bm{G=35}$}\\

$n=240$ & $n=500$ & $n=740$ & $n=1000$  \\
\midrule
\shortstack{$18.1$ \\ $(1.35)$}& \shortstack{$12.4$ \\ $(0.50)$}   & \shortstack{$11.1$ \\ $(0.38)$}  & \shortstack{$10.1$ \\ $(0.25)$} \\ 
\bottomrule
\end{tabular}
\begin{tablenotes}
\footnotesize
\item Notes: 250 repetitions. Standard deviations in parenthesis. DGP as described in Equation \ref{modelmanycov}.
\end{tablenotes}
\end{threeparttable}
\end{table}

\begin{table}[h!]

\centering
\caption{Absolute row sum of $\bm{\kappa}_n$ for $K_n/n=0.300$ - Linear regression model with many continuous controls}
\label{absrowsum3}
\addtolength{\tabcolsep}{+5pt} 
\begin{threeparttable}
\begin{tabular}{lccc}
\hline
\multicolumn{3}{@{}c}{$\bm{G=140}$}\\

$n=250$ & $n=500$ & $n=750$ & $n=1000$  \\
\midrule
\shortstack{$9.36$ \\ $(0.65)$}& \shortstack{$7.66$ \\ $(0.32)$}   & \shortstack{$7.12$ \\ $(0.21)$}  & \shortstack{$6.77$ \\ $(0.17)$} \\ 
\midrule
\multicolumn{3}{@{}c}{$\bm{G=70}$}\\

$n=250$ & $n=500$ & $n=750$ & $n=1000$  \\
\midrule
\shortstack{$17.5$ \\ $(1.10)$}& \shortstack{$13.2$ \\ $(0.67)$}   & \shortstack{$11.9$ \\ $(0.37)$}  & \shortstack{$11.2$ \\ $(0.29)$} \\ 
\midrule
\multicolumn{3}{@{}c}{$\bm{G=35}$}\\

$n=240$ & $n=500$ & $n=740$ & $n=1000$  \\
\midrule
\shortstack{$41.2$ \\ $(2.93)$}& \shortstack{$26.12$ \\ $(1.13)$}   & \shortstack{$22.6$ \\ $(0.77)$}  & \shortstack{$20.7$ \\ $(0.55)$} \\ 
\bottomrule
\end{tabular}
\begin{tablenotes}
\footnotesize
\item Notes: 250 repetitions. Standard deviations in parenthesis. DGP as described in Equation \ref{modelmanycov}.
\end{tablenotes}
\end{threeparttable}
\end{table}
\newpage

\begin{table}[h!]
\centering
\caption{Absolute row sum of $\bm{\kappa}_n$ for $K_n/n=0.400$ - Linear regression model with many continuous controls}
\label{absrowsum4}
\addtolength{\tabcolsep}{+5pt} 
\begin{threeparttable}
\begin{tabular}{lccc}
\hline
\multicolumn{3}{@{}c}{$\bm{G=140}$}\\

$n=250$ & $n=500$ & $n=750$ & $n=1000$  \\
\midrule
\shortstack{$18.36$ \\ $(1.42)$}& \shortstack{$14.5$ \\ $(0.64)$}   & \shortstack{$13.3$ \\ $(0.50)$}  & \shortstack{$12.6$ \\ $(0.36)$} \\ 
\midrule
\multicolumn{3}{@{}c}{$\bm{G=70}$}\\

$n=250$ & $n=500$ & $n=750$ & $n=1000$  \\
\midrule
\shortstack{$37.3$ \\ $(3.26)$}& \shortstack{$26.6$ \\ $(1.15)$}   & \shortstack{$23.8$ \\ $(0.84)$}  & \shortstack{$22.0$ \\ $(0.59)$} \\ 
\midrule
\multicolumn{3}{@{}c}{$\bm{G=35}$}\\

$n=240$ & $n=500$ & $n=740$ & $n=1000$  \\
\midrule
\shortstack{$100$ \\ $(8.30)$}& \shortstack{$59.9$ \\ $(2.64)$}   & \shortstack{$47.6$ \\ $(1.77)$}  & \shortstack{$43.0$ \\ $(1.20)$} \\ 
\bottomrule
\end{tabular}
\begin{tablenotes}
\footnotesize
\item Notes: 250 repetitions. Standard deviations in parenthesis. DGP as described in Equation \ref{modelmanycov}.
\end{tablenotes}
\end{threeparttable}
\end{table}

\begin{table}[h!]

\centering
\caption{Absolute row sum of $\bm{\kappa}_n$ for $K_n/n=0.200$ - Linear regression model with many discrete controls}
\label{absrowsum2discrete}
\addtolength{\tabcolsep}{+5pt} 
\begin{threeparttable}
\begin{tabular}{lccc}
\hline
\multicolumn{3}{@{}c}{$\bm{G=140}$}\\

$n=250$ & $n=500$ & $n=750$ & $n=1000$  \\
\midrule
\shortstack{$6.12$ \\ $(0.50)$}& \shortstack{$4.94$ \\ $(0.28)$}   & \shortstack{$4.53$ \\ $(0.18)$}  & \shortstack{$4.27$ \\ $(0.14)$} \\ 
\midrule
\multicolumn{3}{@{}c}{$\bm{G=70}$}\\

$n=250$ & $n=500$ & $n=750$ & $n=1000$  \\
\midrule
\shortstack{$10.5$ \\ $(0.85)$}& \shortstack{$8.00$ \\ $(0.48)$}   & \shortstack{$7.10$ \\ $(0.31)$}  & \shortstack{$6.62$ \\ $(0.25)$} \\ 
\midrule
\multicolumn{3}{@{}c}{$\bm{G=35}$}\\

$n=240$ & $n=500$ & $n=740$ & $n=1000$  \\
\midrule
\shortstack{$22.9$ \\ $(2.330)$}& \shortstack{$15.0$ \\ $(0.94)$}   & \shortstack{$12.8$ \\ $(0.56)$}  & \shortstack{$11.7$ \\ $(0.51)$} \\ 
\bottomrule
\end{tabular}
\begin{tablenotes}
\footnotesize
\item Notes: 250 repetitions. Standard deviations in parenthesis. DGP as described in Equation \ref{modelmanycov}, with $w_{\ell,gi}=\mathbbm{1}\{\mathcal{N}(0,1,)\geq 1\},\quad \forall \ell, g, i$.
\end{tablenotes}
\end{threeparttable}
\end{table}

\begin{table}[h!]

\centering
\caption{Absolute row sum of $\bm{\kappa}_n$ for $K_n/n=0.300$ - Linear regression model with many discrete controls}
\label{absrowsum3discrete}
\addtolength{\tabcolsep}{+5pt} 
\begin{threeparttable}
\begin{tabular}{lccc}
\hline
\multicolumn{3}{@{}c}{$\bm{G=140}$}\\

$n=250$ & $n=500$ & $n=750$ & $n=1000$  \\
\midrule
\shortstack{$11.2$ \\ $(0.96)$}& \shortstack{$8.88$ \\ $(0.51)$}   & \shortstack{$8.04$ \\ $(0.34)$}  & \shortstack{$7.57$ \\ $(0.28)$} \\ 
\midrule
\multicolumn{3}{@{}c}{$\bm{G=70}$}\\

$n=250$ & $n=500$ & $n=750$ & $n=1000$  \\
\midrule
\shortstack{$20.8$ \\ $(1.62)$}& \shortstack{$15.3$ \\ $(0.83)$}   & \shortstack{$13.4$ \\ $(0.57)$}  & \shortstack{$12.6$ \\ $(0.47)$} \\ 
\midrule
\multicolumn{3}{@{}c}{$\bm{G=35}$}\\

$n=240$ & $n=500$ & $n=740$ & $n=1000$  \\
\midrule
\shortstack{$49.9$ \\ $(4.70)$}& \shortstack{$30.5$ \\ $(1.82)$}   & \shortstack{$25.8$ \\ $(1.19)$}  & \shortstack{$23.4$ \\ $(0.86)$} \\ 
\bottomrule
\end{tabular}
\begin{tablenotes}
\footnotesize
\item Notes: 250 repetitions. Standard deviations in parenthesis. DGP as described in Equation \ref{modelmanycov}, with $w_{\ell,gi}=\mathbbm{1}\{\mathcal{N}(0,1,)\geq 1\},\quad \forall \ell, g, i$.
\end{tablenotes}
\end{threeparttable}
\end{table}

\begin{table}[h!]

\centering
\caption{Absolute row sum of $\bm{\kappa}_n$ for $K_n/n=0.400$ - Linear regression model with many discrete controls}
\label{absrowsum4discrete}
\addtolength{\tabcolsep}{+5pt} 
\begin{threeparttable}
\begin{tabular}{lccc}
\hline
\multicolumn{3}{@{}c}{$\bm{G=140}$}\\

$n=250$ & $n=500$ & $n=750$ & $n=1000$  \\
\midrule
\shortstack{$21.4$ \\ $(1.87)$}& \shortstack{$16.5$ \\ $(0.85)$}   & \shortstack{$14.9$ \\ $(0.68)$}  & \shortstack{$13.9$ \\ $(0.53)$} \\ 
\midrule
\multicolumn{3}{@{}c}{$\bm{G=70}$}\\

$n=250$ & $n=500$ & $n=750$ & $n=1000$  \\
\midrule
\shortstack{$43.5$ \\ $(4.49)$}& \shortstack{$30.3$ \\ $(1.66)$}   & \shortstack{$26.3$ \\ $(1.11)$}  & \shortstack{$24.3$ \\ $(0.88$} \\ 
\midrule
\multicolumn{3}{@{}c}{$\bm{G=35}$}\\

$n=240$ & $n=500$ & $n=740$ & $n=1000$  \\
\midrule
\shortstack{$119.4$ \\ $(11.7)$}& \shortstack{$64.6$ \\ $(4.22)$}   & \shortstack{$53.1$ \\ $(2.47)$}  & \shortstack{$47.5$ \\ $(1.90)$} \\ 
\bottomrule
\end{tabular}
\begin{tablenotes}
\footnotesize
\item Notes: 250 repetitions. Standard deviations in parenthesis. DGP as described in Equation \ref{modelmanycov}, with $w_{\ell,gi}=\mathbbm{1}\{\mathcal{N}(0,1,)\geq 1\},\quad \forall \ell, g, i$.
\end{tablenotes}
\end{threeparttable}
\end{table}

\begin{table}[h!]

\centering
\caption{Absolute row sum of $\bm{\kappa}_n$ for $K_n/n=0.200$ - Semiparametric partially linear model}
\label{absrowsumsemi2}
\addtolength{\tabcolsep}{+5pt} 
\begin{threeparttable}
\begin{tabular}{lcccc}
\hline
\multicolumn{3}{@{}c}{$\bm{G=140}$}\\

$n=250$ & $n=500$ & $n=750$ & $n=1000$\\
\midrule
\shortstack{$19.6$ \\ $(6.01)$}& \shortstack{$18.1$ \\ $(4.86)$}   & \shortstack{$43.3$ \\ $(18.6)$}  & \shortstack{$43.7$ \\ $(17.2)$}\\ 
\midrule
\multicolumn{3}{@{}c}{$\bm{G=70}$}\\

$n=250$ & $n=500$ & $n=750$ & $n=1000$  \\
\midrule
\shortstack{$33.0$ \\ $(11.3)$}& \shortstack{$27.5$ \\ $(8.99)$}   & \shortstack{$63.2$ \\ $(28.8)$}  & \shortstack{$63.1$ \\ $(25.1)$} \\ 
\midrule
\multicolumn{3}{@{}c}{$\bm{G=35}$}\\

$n=240$ & $n=500$ & $n=740$ & $n=1000$  \\
\midrule
\shortstack{$75.1$ \\ $(26.8)$}& \shortstack{$51.8$ \\ $(15.0)$}   & \shortstack{$101.2$ \\ $(41.6)$}  & \shortstack{$97.6$ \\ $(33.0)$} \\ 
\bottomrule
\end{tabular}
\begin{tablenotes}
\footnotesize
\item Notes: 250 repetitions. Standard deviations in parenthesis. DGP as described in Equation \ref{modelpartially}.
\end{tablenotes}
\end{threeparttable}
\end{table}

\begin{table}[h!]

\centering
\caption{Absolute row sum of $\bm{\kappa}_n$ for $K_n/n=0.300$ - Semiparametric partially linear model}
\label{absrowsumsemi3}
\addtolength{\tabcolsep}{+5pt} 
\begin{threeparttable}
\begin{tabular}{lcccc}
\hline
\multicolumn{3}{@{}c}{$\bm{G=140}$}\\

$n=250$ & $n=500$ & $n=750$ & $n=1000$\\
\midrule
\shortstack{$64.9$ \\ $(41.1)$}& \shortstack{$142.5$ \\ $(70.7)$}   & \shortstack{$142.4$ \\ $(63.9)$}  & \shortstack{$56.1$ \\ $(25.0)$}\\ 
\midrule
\multicolumn{3}{@{}c}{$\bm{G=70}$}\\

$n=250$ & $n=500$ & $n=750$ & $n=1000$  \\
\midrule
\shortstack{$108.7$ \\ $(46.9)$}& \shortstack{$210.9$ \\ $(103.9)$}   & \shortstack{$190.8$ \\ $(84.7)$}  & \shortstack{$80.1$ \\ $(34.8)$} \\ 
\midrule
\multicolumn{3}{@{}c}{$\bm{G=35}$}\\

$n=240$ & $n=500$ & $n=740$ & $n=1000$  \\
\midrule
\shortstack{$260.4$ \\ $(108.4)$}& \shortstack{$389.7$ \\ $(176.7)$}   & \shortstack{$355.7$ \\ $(161.1)$}  & \shortstack{$134.4$ \\ $(62.5)$} \\ 
\bottomrule
\end{tabular}
\begin{tablenotes}
\footnotesize
\item Notes: 250 repetitions. Standard deviations in parenthesis. DGP as described in Equation \ref{modelpartially}.
\end{tablenotes}
\end{threeparttable}
\end{table}

\begin{table}[h!]

\centering
\caption{Absolute row sum of $\bm{\kappa}_n$ for $K_n/n=0.400$ - Semiparametric partially linear model}
\label{absrowsumsemi4}
\addtolength{\tabcolsep}{+5pt} 
\begin{threeparttable}
\begin{tabular}{lcccc}
\hline
\multicolumn{3}{@{}c}{$\bm{G=140}$}\\

$n=250$ & $n=500$ & $n=750$ & $n=1000$\\
\midrule
\shortstack{$140.7$ \\ $(56.4)$}& \shortstack{$509.0$ \\ $(349.1)$}   & \shortstack{$139.5$ \\ $(65.2)$}  & \shortstack{$56.1$ \\ $(25.0)$}\\ 
\midrule
\multicolumn{3}{@{}c}{$\bm{G=70}$}\\

$n=250$ & $n=500$ & $n=750$ & $n=1000$  \\
\midrule
\shortstack{$265.2$ \\ $(143.0)$}& \shortstack{$700.4$ \\ $(531.5)$}   & \shortstack{$207.9$ \\ $(97.4)$}  & \shortstack{$79.7$ \\ $(30.4)$} \\ 
\midrule
\multicolumn{3}{@{}c}{$\bm{G=35}$}\\

$n=240$ & $n=500$ & $n=740$ & $n=1000$  \\
\midrule
\shortstack{$703.1$ \\ $(324.2)$}& \shortstack{$1321.6$ \\ $(712.4)$}   & \shortstack{$355.6$ \\ $(164.1)$}  & \shortstack{$130.1$ \\ $(48.9)$} \\ 
\bottomrule
\end{tabular}
\begin{tablenotes}
\footnotesize
\item Notes: 250 repetitions. Standard deviations in parenthesis. DGP as described in Equation \ref{modelpartially}.
\end{tablenotes}
\end{threeparttable}
\end{table}

\begin{table}[h!]

\centering
\caption{Absolute row sum of $\bm{\kappa}_n$ for $K_n/n=0.200$ - Two-way fixed effects panel data regression model}
\label{absrowsumtwoway2}
\addtolength{\tabcolsep}{+5pt} 
\begin{threeparttable}
\begin{tabular}{lcccc}
\hline
\multicolumn{3}{@{}c}{$\bm{G=140}$}\\

$n=250$ & $n=500$ & $n=750$ & $n=1000$\\
\midrule
\shortstack{$3.97$ \\ $(.063)$}& \shortstack{$3.90$ \\ $(.039)$}   & \shortstack{$3.87$ \\ $(.031)$}  & \shortstack{$3.85$ \\ $(.030)$}\\ 
\midrule
\multicolumn{3}{@{}c}{$\bm{G=70}$}\\

$n=250$ & $n=500$ & $n=750$ & $n=1000$  \\
\midrule
\shortstack{$3.92$ \\ $(.040)$}& \shortstack{$3.87$ \\ $(.026)$}   & \shortstack{$3.83$ \\ $(.017)$}  & \shortstack{$3.81$ \\ $(.015)$} \\ 
\midrule
\multicolumn{3}{@{}c}{$\bm{G=35}$}\\

$n=240$ & $n=500$ & $n=740$ & $n=1000$  \\
\midrule
\shortstack{$8.25$ \\ $(1.94)$}& \shortstack{$3.89$ \\ $(.023)$}   & \shortstack{$7.16$ \\ $(1.03)$}  & \shortstack{$3.84$ \\ $(.011)$} \\ 
\bottomrule
\end{tabular}
\begin{tablenotes}
\footnotesize
\item Notes: 250 repetitions. Standard deviations in parenthesis. DGP as described in Equation \ref{twowaydgp}.
\end{tablenotes}
\end{threeparttable}
\end{table}

\begin{table}[h!]

\centering
\caption{Absolute row sum of $\bm{\kappa}_n$ for $K_n/n=0.333$ - Two-way fixed effects panel data regression model}
\label{absrowsumtwoway3}
\addtolength{\tabcolsep}{+5pt} 
\begin{threeparttable}
\begin{tabular}{lcccc}
\hline
\multicolumn{3}{@{}c}{$\bm{G=140}$}\\

$n=255$ & $n=510$ & $n=750$ & $n=1005$\\
\midrule
\shortstack{$11.6$ \\ $(.42)$}& \shortstack{$11.1$ \\ $(.30)$}   & \shortstack{$10.9$ \\ $(.23)$}  & \shortstack{$10.8$ \\ $(.21)$}\\ 
\midrule
\multicolumn{3}{@{}c}{$\bm{G=70}$}\\

$n=240$ & $n=510$ & $n=750$ & $n=990$  \\
\midrule
\shortstack{$10.6$ \\ $(.18)$}& \shortstack{$10.2$ \\ $(.11)$}   & \shortstack{$10.0$ \\ $(.098)$}  & \shortstack{$9.90$ \\ $(.074)$} \\ 
\midrule
\multicolumn{3}{@{}c}{$\bm{G=35}$}\\

$n=240$ & $n=480$ & $n=720$ & $n=960$  \\
\midrule
\shortstack{$10.7$ \\ $(.17)$}& \shortstack{$10.2$ \\ $(.082)$}   & \shortstack{$9.91$ \\ $(.062)$}  & \shortstack{$9.77$ \\ $(.047)$} \\ 
\bottomrule
\end{tabular}
\begin{tablenotes}
\footnotesize
\item Notes: 250 repetitions. Standard deviations in parenthesis. DGP as described in Equation \ref{twowaydgp}.
\end{tablenotes}
\end{threeparttable}
\end{table}

\clearpage

\renewcommand{\arraystretch}{1.3}
\begin{table}[h!]
\centering
\caption{Empirical illustration - Effect of Abortion on Crime}
\label{empirical findings}

\addtolength{\tabcolsep}{-0.8pt} 
\begin{threeparttable}
\begin{tabular}{lcccccccc}
\hline
\hline
&&& $\hat{\beta}$ & \multicolumn{2}{c}{LZ} & \multicolumn{2}{c}{Robust} \\
 \cmidrule(l{1pt}r{1pt}){5-6} \cmidrule(l{1pt}r{1pt}){7-9} 
&&&&Std. Error&p-value&Std. Error &p-value& \\ \toprule
\multicolumn{3}{@{}l}{\textbf{Violent crime}}\\

Baseline&&&$-0.135$&$0.0422$&$0.0013$&$0.0448$&$0.0025$\\
Many controls&&&$-0.266$&$0.0842$&$0.0016$&$0.1473$&$0.0718$\\
\midrule
\multicolumn{3}{@{}l}{\textbf{Property crime}}\\

Baseline&&&$-0.093$&$0.0146$&$<0.00001$&$0.0149$&$<0.00001$\\
Many controls&&&$-0.135$&$0.0254$&$<0.00001$&$0.0408$&$0.0009$\\
\midrule
\multicolumn{3}{@{}l}{\textbf{Murder}}\\

Baseline&&&$-0.134$&$0.0536$&$0.0126$&$0.0551$&$0.0154$\\
Many controls&&&$-0.197$&$0.1498$&$0.1848$&$0.2117$&$0.3513$\\

\midrule
\bottomrule
\end{tabular}
\begin{tablenotes}
\footnotesize 
\item The rows labeled ``Baseline estimates" give estimates for the original model in \citet*{dl2001} as in (\ref{DL}). The rows labeled "Many controls" give estimates for the high-dimensional model in (\ref{DLhighdimensional}) that includes a broader set of controls. Columns under the label ``LZ" report clustered standard errors and relative p-values computed with the traditional variance formula by \citet*{liangzeger}. Columns under the label ``Robust" report standard errors and relative p-values computed with the cluster-robust variance estimator proposed in this paper. Clustering is at the state level.
\item Data source: \citet{belloni2014rev}.
\end{tablenotes}
\end{threeparttable}
\end{table}

\clearpage

\renewcommand*\appendixpagename{Appendix}
\renewcommand*\appendixtocname{Appendix}

\small
\setstretch{1.2}
\setcounter{section}{0}
\setcounter{equation}{0}
\setcounter{theorem}{0}
\setcounter{coro}{0}
\setcounter{assum}{0}
\appendix
\appendixpage

\setcounter{subsection}{0}
This appendix is organized as follows. Section \ref{sectionA} presents the assumptions and the variance estimators studied in this paper for the fully general case that allows for misspecification bias in the model. Section \ref{sectionB} presents the main results of the paper under the setup described in Section \ref{sectionA}. Section \ref{sectionC} presents the technical lemmas needed to establish the main results of the paper. Section \ref{sectionD} provides the proofs for the main results of the paper. Section \ref{sectionE} provides the proofs for the technical lemmas. Section \ref{sectionF} presents an extension of the variance estimators studied in the paper which allows to impose within-cluster zero restrictions on the variance-covariance matrix of the errors.
\section{Setup - general case}\label{sectionA}

\newtheorem{innercustomassum}{Assumption}
\newenvironment{customassum}[1]
  {\renewcommand\theinnercustomassum{#1}\innercustomassum}
  {\endinnercustomassum}

\subsection{Assumptions}
Suppose that $\{(y_{i,n},\mathbf{x}_{i,n}',\mathbf{w}_{i,n}') : 1 \leq i \leq n\}$ is generated by 
\begin{equation}
y_{i,n}=\bm{\beta}'\mathbf{x}_{i,n} + \bm{\gamma}_n'\mathbf{w}_{i,n} + u_{i,n},\qquad i=1,\dots, n,
\end{equation}for which $\mathcal{W}_n$ is a collection of random variables such that $\mathbb{E}[\mathbf{w}_{i,n}|\mathcal{W}_n]=\mathbf{w}_{i,n}$, and we set $\mathcal{X}_{n}=(\mathbf{x}_{1,n},\dots,\mathbf{x}_{n,n})$. We define the following quantities:
\begin{equation}
\begin{split}
&\varrho_n=\frac{1}{n}\sum_{i=1}^{n}\mathbb{E}[R_{i,n}^2], \qquad R_{i,n}=\mathbb{E}[u_{i,n}|\mathcal{X}_n,\mathcal{W}_n],\\
&\rho_n=\frac{1}{n}\sum_{i=1}^{n}\mathbb{E}[r_{i,n}^2], \qquad r_{i,n}=\mathbb{E}[u_{i,n}|\mathcal{W}_n],\\
&\chi_n=\frac{1}{n}\sum_{i=1}^{n}\mathbb{E}[\|\mathbf{Q}_{i,n}\|^2], \qquad \mathbf{Q}_{i,n}=\mathbb{E}[\mathbf{v}_{i,n}|\mathcal{W}_n],\\
&\bm{\hat{\Gamma}}_n=\sum_{i=1}^n\mathbf{\hat{v}}_{i,n}\mathbf{\hat{v}}_{i,n}'/n, \qquad \bm{\Sigma}_n=\mathbb{V}[\frac{1}{\sqrt{n}}\sum_{i=1}^{n}\mathbf{\hat{v}}_{i,n}U_{i,n}|\mathcal{X}_n,\mathcal{W}_n],
\end{split}
\end{equation}
where $\mathbf{v}_{i,n}=\mathbf{x}_{i,n}-(\sum_{j=1}^{n}\mathbb{E}[\mathbf{x}_{j,n}\mathbf{w}_{j,n}'])(\sum_{j=1}^{n}\mathbb{E}[\mathbf{w}_{j,n}\mathbf{w}_{j,n}'])^{-1}\mathbf{w}_{i,n}$ is the population counterpart of $\mathbf{\hat{v}}_{i,n}$. Also, letting $\lambda_{\text{min}}(\cdot)$ denote the minimum eigenvalue of its argument, define
\begin{equation}
\mathcal{C}_n=\max_{1\leq i \leq n}{\{\mathbb{E}[U_{i,n}^4|\mathcal{X}_n,\mathcal{W}_n]+\mathbb{E}[\|\mathbf{V}_{i,n}\|^4 |\mathcal{W}_n]+1/\mathbb{E}[U_{i,n}^2|\mathcal{X}_n,\mathcal{W}_n]}\}+1/\lambda_{\text{min}}(\mathbb{E}[\bm{\tilde{\Gamma}}_n|\mathcal{W}_n]),
\end{equation}
where $U_{i,n}=y_{i.n}-\mathbb{E}[y_{i,n}|\mathcal{X}_n, \mathcal{W}_n]$, $\mathbf{V}_{i,n}=\mathbf{x}_{i,n}-\mathbb{E}[\mathbf{x}_{i,n}|\mathcal{W}_n]$, $\mathbf{\tilde{\Gamma}}_n=\sum_{i=1}^{n}\mathbf{\tilde{V}}_{i,n}\mathbf{\tilde{V}}_{i,n}'/n$ and $\mathbf{\tilde{V}}_{i,n}=\sum_{j=1}^{n}M_{ij,n}\mathbf{V}_{i,n}$.\\
\par We impose the following three assumptions:
\begin{customassum}{1*}
$\max_{1\leq g\leq G_n}\#\mathcal{T}_{g,n}=O(1)$, where $\#\mathcal{T}_{g,n}$ is the cardinality of $\mathcal{T}_{g,n}$ and where $\{\mathcal{T}_{g,n}:1\leq g\leq G_n\}$ is a partition of \{1,\dots,n\} such that $\{(U_{i,n},\mathbf{x}'_{i,n}):i\in\mathcal{T}_{g,n}\}$ are independent over g conditional on $\mathcal{W}_{n}$.
\end{customassum}
\begin{customassum}{2*}
 $\mathbb{P}[\lambda_{\text{min}}(\sum_{i=1}^{n}\mathbf{w}_{i,n}\mathbf{w}_{i,n}')>0]\rightarrow 1$, $\limsup_{n\rightarrow\infty}K_n/n < 1$, $\mathcal{C}_n=O_p(1)$ and $\bm{\Sigma}_n^{-1}=O_p(1)$
\end{customassum}
\begin{customassum}{3*}
$\chi_n=O(1)$, $\varrho_n+n(\varrho_n-\rho_n)+n\chi_n\varrho_n=o(1)$, and $\max_{1\leq i \leq n}\|\mathbf{\hat{v}}_{i,n}\|/\sqrt{n}=o_p(1)$.
\end{customassum}

The only difference with the simplified set of assumptions presented in Section 3 of this paper is that we now allow for misspecification bias, i.e. $\mathbb{E}[u_{i}|\mathcal{X}_n, \mathcal{W}_n]\neq0$. In particular, Assumption 3* now also includes conditions on $\varrho_n$ and $\rho_n$, which are requirements on the quality of the linear approximation for the conditional expectations $\mathbb{E}[y_{i,n}|\mathcal{X}_n, \mathcal{W}_n]$ and $\mathbb{E}[y_{i,n}|\mathcal{W}_n]$, respectively. The misspecification bias is required to vanish asymptotically, thus ruling out the presence of lagged outcomes in the model. Notice that when no misspecification bias is present one gets $\varrho_n=\rho_n=0$ and this set of assumptions reduces to the one presented in Section 3 of the paper.

\subsection{Variance estimators}

Let $\bm{\Omega}_{U,n}=\mathbb{E}[\mathbf{U}_n\mathbf{U}_n'|\mathcal{X}_n,\mathcal{W}_n]$ be the (conditional) variance-covariance matrix of the errors $\mathbf{U}_{n}=(U_{1,n}, \dots, U_{n,n})'$ and $L_n=\sum_{g=1}^{G_n}(\#\mathcal{T}_{g,n})^2$ the number of non-zero elements contained in it.
We define a general class of cluster-robust estimators for $\bm{\Sigma}_n$ of the form
\begin{equation}
\bm{\hat{\Sigma}}_n^{}(\bm{\kappa}_n)=\frac{1}{n}\sum_{g_1=1}^{G_n}\sum_{g_2=1}^{G_n}\sum_{i_1,j_1\in\mathcal{T}_{g_1,n}}\sum_{i_2,j_2\in\mathcal{T}_{g_2,n}}\kappa_{g_1, g_2, i_1,j_1,i_2,j_2,n}\mathbf{\hat{v}}_{i_1,n}\mathbf{\hat{v}}'_{j_1,n}\hat{u}_{i_2,n}\hat{u}_{j_2,n},
\end{equation}
where $\kappa_{g_1,g_2,i_1,j_1,i_2.j_2,n}$ is an entry of the $L_n\times L_n$ symmetric matrix $\bm{\kappa}_{n}=\bm{\kappa}_{n}(\mathbf{w}_{1,n}, \dots, \mathbf{w}_{1,n})$.\footnote{In particular, $\kappa_{g_1,g_2,i_1,j_1,i_2.j_2,n}$ corresponds to the $(h(g_1,i_1,j_1), h(g_2,i_2,j_2))$ entry of $\bm{\kappa}_{n}$, where $h(g,i,j)=[\sum_{k=0}^{(g-1)}(\#\mathcal{T}_{k,n})^2+(\#\mathcal{T}_{g,n})(i-1) +j]$ and we adopt the convention that $\#\mathcal{T}_{0,n}=0$.}

Furthermore, define
\begin{equation}\label{kronecker}
\bm{\kappa}^{\texttt{CR}}_n=(\mathbf{S}_n'(\mathbf{M}_n\otimes\mathbf{M}_n) \mathbf{S}_n)^{-1},
\end{equation}
where $\otimes$ denotes the Kronecker product and $\mathbf{S}_n$ is the $n^2 \times L_n$ selection matrix with full column rank such that $\mathbf{S}_n'\text{vec}(\bm{\Omega}_{U,n})$ is the $L_n\times 1$ vector containing the non-zero elements of $\bm{\Omega}_{U,n}$. Our proposed cluster-robust estimator is then defined as
\begin{equation}
\bm{\hat{\Sigma}}_n^{\texttt{CR}}\equiv \bm{\hat{\Sigma}}^{\texttt{}}(\bm{\kappa}^{\texttt{CR}}_n)=\frac{1}{n}\sum_{g_1=1}^{G_n}\sum_{g_2=1}^{G_n}\sum_{i_1,j_1\in\mathcal{T}_{g_1,n}}\sum_{i_2,j_2\in\mathcal{T}_{g_2,n}}\kappa_{g_1, g_2, i_1,j_1,i_2,j_2,n}^{\texttt{CR}}\mathbf{\hat{v}}_{i_1,n}\mathbf{\hat{v}}'_{j_1,n}\hat{u}_{i_2,n}\hat{u}_{j_2,n}.
\end{equation}

\section{Main results - general case}\label{sectionB}
\newtheorem{innercustomthm}{Theorem}
\newenvironment{customthm}[1]
  {\renewcommand\theinnercustomthm{#1}\innercustomthm}
  {\endinnercustomthm}

\newtheorem{innercustomcoro}{Corollary}
\newenvironment{customcoro}[1]
  {\renewcommand\theinnercustomcoro{#1}\innercustomcoro}
  {\endinnercustomcoro}
In this section we present the generalisation of the main results of the paper to the case of potential misspecification bias in the model.

The first theorem establishes asymptotic normality of the OLS estimator for $\bm{\beta}_n$.

\begin{customthm}{1*}
 Suppose Assumptions 1*-3*  hold. Then,
\begin{equation}
\bm{\Omega}_n^{-1/2}\sqrt[]{n}(\bm{\hat{\beta}}_n-\bm{\beta})\overset{d}{\to}\mathcal{N}(0,\mathbf{I}_d), \qquad \bm{\Omega}_n=\bm{\hat{\Gamma}}^{-1}_n\bm{\Sigma}_n\bm{\hat{\Gamma}}^{-1}_n,
\end{equation}
where $\bm{\Sigma}_n=\frac{1}{n}\sum_{g=1}^{G_n}\sum_{i,j\in\mathcal{T}_{g,n}}\mathbf{\hat{v}}_{i,n}\mathbf{\hat{v}}'_{j,n}\mathbb{E}[U_{i,n}U_{j,n}|\mathcal{X}_n,\mathcal{W}_n]$.
\end{customthm}

The second theorem provides an asymptotic representation for the general class of variance estimators defined in (4).

\begin{customthm}{2*} Suppose Assumptions 1*-3* hold.\\ If $\norm{\bm{\kappa}_{n}}_\infty=\max_{(g_1,i_1,j_1)}\sum_{g_2=1}^{G_n}\sum_{i_2,j_2\in\mathcal{T}_{g_2,n}}|\kappa_{g_1, g_2, i_1,j_1,i_2,j_2,n}|= O_p(1)$, then
\begin{equation}\begin{split}
\bm{\hat{\Sigma}}_n^{\textup{\texttt{}}}(\bm{\kappa}_n)=\\ \frac{1}{n}\sum_{g_1=1}^{G_n}\sum_{g_2=1}^{G_n}\sum_{g_3=1}^{G_n}\sum_{i_1,j_1\in\mathcal{T}_{g_1,n}}\sum_{i_2,j_2\in\mathcal{T}_{g_2,n}}\sum_{i_3,j_3\in\mathcal{T}_{g_3,n}}&\kappa_{g_1, g_2, i_1,j_1,i_2,j_2,n}\mathbf{\hat{v}}_{i_1,n}\mathbf{\hat{v}}'_{j_1,n}M_{i_2j_3,n}M_{j_2i_3,n}\mathbb{E}[U_{i_3,n}U_{j_3,n}|\mathcal{X}_n,\mathcal{W}_n]\\
&+o_p(1).
\end{split}
\end{equation}
\end{customthm}

Corollary 1* characterizes the asymptotic limit of LZ's estimator.

\begin{customcoro}{1*} Suppose the assumptions of Theorem 2* hold. Then,
\begin{equation*}
\bm{\hat{\Sigma}}_n^{\textup{\texttt{LZ}}}=\frac{1}{n}\sum_{g_1=1}^{G_n}\sum_{g_2=1}^{G_n}\sum_{i_1,j_1\in\mathcal{T}_{g_1,n}}\sum_{i_2,j_2\in\mathcal{T}_{g_2,n}}\mathbf{\hat{v}}_{i_1,n}\mathbf{\hat{v}}'_{j_1,n}M_{i_1j_2,n}M_{j_1i_2,n}\mathbb{E}[U_{i_2,n}U_{j_2,n}|\mathcal{X}_n,\mathcal{W}_n] + o_p(1).
\end{equation*}
\end{customcoro}

The third theorem establishes consistency of our proposed estimator.

\begin{customthm}{3*} 
Suppose Assumptions 1*-3* hold.\\ If $\mathbb{P}[\lambda_{\text{min}}(\mathbf{S}_n'(\mathbf{M}_n\otimes\mathbf{M}_n) \mathbf{S}_n)>0]\to1$ and $\norm{\bm{\kappa}_{n}^{\textup{\texttt{CR}}}}_\infty= O_p(1)$, then
\begin{equation}
\bm{\hat{\Sigma}}^{\textup{\texttt{CR}}}_n=\bm{\Sigma}_n +o_p(1).
\end{equation}
\end{customthm}

Finally, the fourth theorem provides sufficient conditions for consistency of LZ's estimator. For this purpose, define $\mathbf{w}^{*}_{i,n}=\mathbf{w}_{i,n}\bm{\hat{\Sigma}}_{\mathbf{w},n}^{-1/2}$, where $\bm{\hat{\Sigma}}_{\mathbf{w},n}^{1/2}$ is the unique symmetric positive definite $K_n\times K_n$ matrix such that $\bm{\hat{\Sigma}}_{\mathbf{w},n}^{1/2}\bm{\hat{\Sigma}}_{\mathbf{w},n}^{1/2}=\frac{1}{n}\sum_{i=1}^{n}\mathbf{w}_{i,n}\mathbf{w}_{i,n}'$.

\begin{customthm}{4*}
Suppose Assumptions 1*-3* hold and that $\max_{i,j}\mathbb{E}[w_{ij,n}^{*2}]=O(1)$.
If $K_{n}^2/n\to0$, then
\begin{equation}\label{convergencewhitappend}
\bm{\hat{\Sigma}}^{\textup{\texttt{LZ}}}_n=\bm{\Sigma}_n +o_p(1).
\end{equation}
Moreover, if $\mathbb{E}[U^2_{i,n}|\mathcal{X}_n, \mathcal{W}_n]=\sigma_{n}^2$ $\forall i$, and $\mathbb{E}[U_{i,n}U_{j,n}|\mathcal{X}_n, \mathcal{W}_n]=0$ $\forall i\neq j$, then (\ref{convergencewhitappend}) holds under $K_n/n\to0$.
\end{customthm}

\section{Technical Lemmas}\label{sectionC}
Here we present the technical lemmas needed to establish the main results of the paper.\footnote{Throughout the Technical Lemmas we adopt the notational convention $\sum_{(g,i,j)}\equiv\sum_{g=1}^{G_n}\sum_{i,j\in\mathcal{T}_{g,n}}$}\\

\par The first lemma can be used to approximate $\bm{\hat{\Sigma}}_n(\bm{\kappa}_n)$ by means of $\bm{\tilde{\Sigma}}_n(\bm{\kappa}_n)$, where
\begin{align*}
&\bm{\hat{\Sigma}}_n(\bm{\kappa}_n)=\frac{1}{n}\sum_{(g_1,i_1,j_1)}\sum_{(g_2,i_2,j_2)}\kappa_{g_1, g_2, i_1,j_1,i_2,j_2,n}\mathbf{\hat{v}}_{i_1,n}\mathbf{\hat{v}}'_{j_1,n}\hat{u}_{i_2,n}\hat{u}_{j_2,n},\\
&\bm{\tilde{\Sigma}}_n(\bm{\kappa}_n)=\frac{1}{n}\sum_{(g_1,i_1,j_1)}\sum_{(g_2,i_2,j_2)}\kappa_{g_1, g_2, i_1,j_1,i_2,j_2,n}\mathbf{\hat{v}}_{i_1,n}\mathbf{\hat{v}}'_{j_1,n}\tilde{U}_{i_2,n}\tilde{U}_{j_2,n}, \qquad \tilde{U}_{i,n}=\sum_{j=1}^{n}M_{ij,n}U_{j,n}
\end{align*}
\begin{lemma}
Suppose Assumptions 1*-3* hold. If $\norm{\bm{\kappa}_{n}}_\infty= O_p(1)$, then 
\begin{equation*} \bm{\hat{\Sigma}}_n(\bm{\kappa}_n)=\mathbb{E}[\bm{\tilde{\Sigma}}_n(\bm{\kappa}_n)|\mathcal{X}_n, \mathcal{W}_n] + o_p(1).\end{equation*}
\end{lemma}

\par The second lemma can be combined with Lemma 1 to show consistency of $\bm{\hat{\Sigma}}_n(\bm{\kappa}_n)$ under a high-level condition.
\begin{lemma}
Suppose Assumption 2* holds. If
\begin{equation*}\begin{split}
\max_{(g_1,i_1,j_1)}\Big\{\Big| \sum_{(g_2, i_2, j_2)}\kappa_{g_1,g_2,i_1,j_1,i_2,j_2,n}M_{i_1j_2,n}M_{j_1i_2,n}-1\Big|+\sum_{(g_3, i_3, j_3)\neq(g_1,i_1,j_1)}\Big| \sum_{(g_2, i_2, j_2)}\kappa_{g_1,g_2,i_1,j_1,i_2,j_2,n}M_{i_3,j_2,n}M_{j_3,i_2,n}\Big| \Big\} \\
= o_p(1),\end{split}
\end{equation*}
then $\mathbb{E}[\bm{\tilde{\Sigma}}_n(\bm{\kappa}_n)|\mathcal{X}_n, \mathcal{W}_n]= \bm{\Sigma}_n + o_p(1)$.
\end{lemma}
The third lemma gives sufficient conditions for the condition of Lemma 2 for our proposed estimator $\bm{\hat{\Sigma}}^{\texttt{CR}}_n$.
\begin{lemma}
Suppose Assumption 2* holds.
If  $\mathbb{P}[\lambda_{\text{min}}(\mathbf{S}_n'(\mathbf{M}_n\otimes\mathbf{M}_n) \mathbf{S}_n)>0]\to1$, then
\begin{equation*}
\mathbb{E}[\bm{\tilde{\Sigma}}_n(\bm{\kappa}_n^{\textup{\texttt{CR}}})|\mathcal{X}_n, \mathcal{W}_n]=\bm{\Sigma}_n + o_p(1).
\end{equation*}
with $\bm{\kappa}_n^{\textup{\texttt{CR}}}=(\mathbf{S}_n'(\mathbf{M}_n\otimes\mathbf{M}_n) \mathbf{S}_n)^{-1}$.
\end{lemma}
The fourth lemma finds sufficient conditions for the condition of Lemma 2 for LZ's estimator.
\begin{lemma}
Suppose Assumption 2* holds and $\bm{\kappa}_n=\mathbf{I}_{L_n}$. Also define $\mathbf{w}^{*}_{i,n}=\mathbf{w}_{i,n}\bm{\hat{\Sigma}}_{\mathbf{w},n}^{-1/2}$, where $\bm{\hat{\Sigma}}_{\mathbf{w},n}^{1/2}$ is the unique symmetric positive definite $K_n\times K_n$ matrix such that $\bm{\hat{\Sigma}}_{\mathbf{w},n}^{1/2}\bm{\hat{\Sigma}}_{\mathbf{w},n}^{1/2}=\frac{1}{n}\sum_{i=1}^{n}\mathbf{w}_{n}\mathbf{w}_{n}'$. If  $\max_{i,j}\mathbb{E}[w_{ij,n}^{*2}]=O(1)$ and $K_n=o(n^{1/2})$, then
\begin{equation*}
\mathbb{E}[\bm{\tilde{\Sigma}}_n(\bm{\kappa}_n)|\mathcal{X}_n, \mathcal{W}_n]=\bm{\Sigma}_n + o_p(1).
\end{equation*}

\end{lemma}

Finally, the fifth lemma establishes sufficient conditions for the condition of Lemma 2 for LZ's estimator for the special case of homoskedastic errors.

\begin{lemma}
Suppose Assumption 2* holds and $\bm{\kappa}_n=\mathbf{I}_{L_n}$. If  $\max_{i,j}\mathbb{E}[w_{ij,n}^{*2}]=O(1)$, $K_n=o(n)$, $\mathbb{E}[U_{i,n}^2|\mathcal{X}_n, \mathcal{W}_n]=\sigma^2_n \quad \forall i$ and $\mathbb{E}[U_{i,n}U_{j,n}|\mathcal{X}_n, \mathcal{W}_n]=0\quad \forall i\neq j$, then
\begin{equation*}
\mathbb{E}[\bm{\tilde{\Sigma}}_n(\bm{\kappa}_n)|\mathcal{X}_n, \mathcal{W}_n]=\bm{\Sigma}_n + o_p(1).
\end{equation*}

\end{lemma}

\section{Proof of Main Results}\label{sectionD}
Theorem 1* follows from Lemma SA-1 and Lemma SA-2 in \citet{cjn}, combined with the fact that $\bm{\Sigma}^{-1}_n=O_p(1)$ in Assumption 2*. Theorem 2* follows from Theorem 1* combined with Lemma 1. Theorem 3* follows from Theorem 2* combined with Lemma 2 and 3. Theorem 4* follows from Theorem 2* combined with Lemma 2, 4 and 5.

\section{Proofs of Technical Lemmas}\label{sectionE}
\setlength{\mathindent}{0pt}
Here we provide the proofs for the technical lemmas. To simplify notation, throughout the proofs we assume $d=1$ without loss of generality.
\subsection{Proof of Lemma 1}
It suffices to show that $\hat{\Sigma}_n(\bm{\kappa}_n) =\tilde{\Sigma}_n(\bm{\kappa}_n) + o_p(1)$ and that $\tilde{\Sigma}_n(\bm{\kappa}_n)=\mathbb{E}[\tilde{\Sigma}_n(\bm{\kappa}_n)|\mathcal{X}_n,\mathcal{W}_n] + o_p(1)$.\\
First,\begin{displaymath}
\tilde{\Sigma}_n(\bm{\kappa}_n)=\frac{1}{n}\sum_{1\leq i\leq G_n}c_{ii,n} + \frac{2}{n}\sum_{1\leq i,j\leq G_n,i<j}c_{ij,n},
\end{displaymath}
\begin{displaymath}
c_{ij,n}=\sum_{s\in\mathcal{T}_i,t\in\mathcal{T}_j}\sum_{(g_1,i_1,j_1)}\sum_{(g_2,i_2,j_2)}\kappa_{g_1, g_2, i_1,j_1,i_2,j_2,n}\hat{v}_{i_1,n}\hat{v}_{j_1,n}M_{i_2 s,n}M_{j_2 t,n}U_{s,n}U_{t,n},
\end{displaymath}
where $\sum_{1\leq i,j\leq G_n}\mathbb{V}[c_{ij,n}|\mathcal{X}_n\mathcal{W}_n]=o_p(n^2)$ because
\begin{align*}
\mathbb{V}[c_{ij,n}|\mathcal{X}_n,\mathcal{W}_n] & \leq (\#\mathcal{T}_{i,n})(\#\mathcal{T}_{j,n})\sum_{s\in\mathcal{T}_{i,n},t\in\mathcal{T}_{j,n}}(\sum_{(g_1,i_1,j_1)}\sum_{(g_2,i_2,j_2)}\kappa_{g_1, g_2, i_1,j_1,i_2,j_2}\hat{v}_{i_1,n}\hat{v}_{j_1,n}M_{i_2 s,n}M_{j_2 t,n})^2\mathbb{V}[U_{s,n}U_{t,n}|\mathcal{X}_n\mathcal{W}_n]\\
&\leq\mathcal{C}^2_{\mathcal{T},n}\mathcal{C}_{U,n}\sum_{s\in\mathcal{T}_{i,n},t\in\mathcal{T}_{j,n}}(\sum_{(g_1,i_1,j_1)}\sum_{(g_2,i_2,j_2)}\kappa_{g_1, g_2, i_1,j_1,i_2,j_2,n}\hat{v}_{i_1,n}\hat{v}_{j_1,n}M_{i_2s,n}M_{j_2t,n})^2\\
& \leq\mathcal{C}^2_{\mathcal{T},n}\mathcal{C}_{U,n}\sum_{1\leq s, t\leq n}(\sum_{(g_1,i_1,j_1)}\sum_{(g_2,i_2,j_2)}\kappa_{g_1, g_2, i_1,j_1,i_2,j_2,n}\hat{v}_{i_1,n}\hat{v}_{j_1,n}M_{i_2 s,n}M_{j_2t,n})^2\\
& 
=\mathcal{C}^2_{\mathcal{T},n}\mathcal{C}_{U,n}\sum_{1\leq s, t\leq n}\sum_{(g_1,i_1,j_1)}\sum_{(g_2,i_2,j_2)}\sum_{(g_3,i_3,j_3)}\sum_{(g_4,i_4,j_4)}\\&\kappa_{g_1, g_2, i_1,j_1,i_2,j_2,n}\kappa_{g_3, g_4, i_3,j_3,i_4,j_4,n}\hat{v}_{i_1,n}\hat{v}_{j_1,n}\hat{v}_{i_3,n}\hat{v}_{j_3,n}M_{i_2s,n}M_{j_2t,n}M_{i_4s,n}M_{j_4t,n}\\
& 
=\mathcal{C}^2_{\mathcal{T},n}\mathcal{C}_{U,n}\sum_{(g_1,i_1,j_1)}\sum_{(g_2,i_2,j_2)}\sum_{(g_3,i_3,j_3)}\sum_{(g_4,i_4,j_4)}\\&\kappa_{g_1, g_2, i_1,j_1,i_2,j_2,n}\kappa_{g_3, g_4, i_3,j_3,i_4,j_4,n}\hat{v}_{i_1,n}\hat{v}_{j_1,n}\hat{v}_{i_3,n}\hat{v}_{j_3,n}M_{i_2 i_4,n}M_{j_2 j_4,n}\\
& 
\leq\mathcal{C}^2_{\mathcal{T},n}\mathcal{C}_{U,n}\sum_{(g_1,i_1,j_1)}\sum_{(g_2,i_2,j_2)}\sum_{(g_3,i_3,j_3)}\sum_{(g_4,i_4,j_4)}\\&|\kappa_{g_1, g_2, i_1,j_1,i_2,j_2,n}||\kappa_{g_3, g_4, i_3,j_3,i_4,j_4,n}||\hat{v}_{i_1,n}||\hat{v}_{j_1,n}||\hat{v}_{i_3,n}||\hat{v}_{j_3,n}||M_{i_2 i_4,n}||M_{j_2 j_4,n}|,
\end{align*}
where $\mathcal{C}_{\mathcal{T},n}=\max_{1\leq i\leq G_n}\#(\mathcal{T}_{i,n})$, $\mathcal{C}_{U,n}=1+\max_{1\leq i \leq n}\mathbb{E}[U_{i,n}^4|\mathcal{X}_n, \mathcal{W}_n]$, and
\begin{flalign*}
&\sum_{(g_1,i_1,j_1)}\sum_{(g_2,i_2,j_2)}\sum_{(g_3,i_3,j_3)}\sum_{(g_4,i_4,j_4)}|\kappa_{g_1, g_2, i_1,j_1,i_2,j_2,n}||\kappa_{g_3, g_4, i_3,j_3,i_4,j_4,n}||\hat{v}_{i_1,n}||\hat{v}_{j_1,n}||\hat{v}_{i_3,n}||\hat{v}_{j_3,n}||M_{i_2 i_4,n}||M_{j_2 j_4,n}|\\
&\leq (\max_{1\leq i\leq n} |\hat{v}_{i,n}|)^2\sum_{(g_1,i_1,j_1)}\sum_{(g_2,i_2,j_2)}\sum_{(g_3,i_3,j_3)}\sum_{(g_4,i_4,j_4)}|\kappa_{g_1, g_2, i_1,j_1,i_2,j_2,n}||\kappa_{g_3, g_4, i_3,j_3,i_4,j_4,n}||\hat{v}_{i_3,n}||\hat{v}_{j_3,n}||M_{i_2 i_4,n}||M_{j_2 j_4,n}|\\
&\leq (\max_{1\leq i\leq n} |\hat{v}_{i,n}|)^2\norm{\bm{\kappa}_{n}}_\infty\sum_{(g_2,i_2,j_2)}\sum_{(g_3,i_3,j_3)}\sum_{(g_4,i_4,j_4)}|\kappa_{g_3, g_4, i_3,j_3,i_4,j_4,n}||\hat{v}_{i_3,n}||\hat{v}_{j_3,n}||M_{i_2 i_4,n}||M_{j_2 j_4,n}|\\
&\leq (\max_{1\leq i\leq n} |\hat{v}_{i,n}|)^2\norm{\bm{\kappa}_{n}}_\infty\mathcal{C}_{\mathcal{T},n}\sum_{(g_3,i_3,j_3)}\sum_{(g_4,i_4,j_4)}|\kappa_{g_3, g_4, i_3,j_3,i_4,j_4,n}||\hat{v}_{i_3,n}||\hat{v}_{j_3,n}|\\
&\leq (\max_{1\leq i\leq n} |\hat{v}_{i,n}|)^2\norm{\bm{\kappa}_{n}}_{\infty}^{2}\mathcal{C}_{\mathcal{T},n}\sum_{(g_3,i_3,j_3)}|\hat{v}_{i_3,n}||\hat{v}_{j_3,n}|\\
&\leq n^2 (\frac{\max_{1\leq i\leq n} |\hat{v}_i|}{\sqrt{n}})^2\norm{\bm{\kappa}_{n}}_{\infty}^{2}\mathcal{C}_{\mathcal{T},n}^2(\frac{1}{n}\sum_{1\leq i\leq n}\hat{v}_{i,n}^2)=o_p(n^2),
\end{flalign*}

where the third inequality uses

\begin{align*}
\sum_{g_2=1}^{G_n}\sum_{i_2,j_2\in\mathcal{T}_{g_2,n}}|M_{i_2 i_4,n}||M_{j_2 j_4,n}|&\leq \sqrt{\big(\sum_{g_2=1}^{G_n}\sum_{i_2,j_2)\in\mathcal{T}_{g_2,n}}M_{i_2 i_4,n}^2\big)(\sum_{g_2=1}^{G_n}\sum_{(i_2,j_2)\in\mathcal{V}_{g_2,n}}M_{j_2 j_4,n}^2)}\\
&\leq \sqrt{\big(\mathcal{C}_{\mathcal{T},n}\sum_{k=1}^{n}M_{k i_4,n}^2\big)\big(\mathcal{C}_{\mathcal{T},n}\sum_{l=1}^{n}M_{l j_4,n}^2\big)}\\
&=\mathcal{C}_{\mathcal{T},n}\sqrt[]{M_{i_4 i_4,n}M_{j_4 j_4,n}}\\
&\leq \mathcal{C}_{\mathcal{T},n}, \end{align*}
and the last inequality similarly uses\footnote{We also make use of the bound $\frac{1}{n}\sum_{i=1}^{n}\hat{v}_{i}^2=O_p(1)$, as shown in Lemma SA-1 of \citeauthor{cjn} (\citeyear{cjn}, Supplemental Appendix).}
\begin{align*}
\sum_{g_3=1}^{G}\sum_{i_3,j_3\in\mathcal{T}_{g_2,n}}|\hat{v}_{i_3,n}||\hat{v}_{j_3,n}|&\leq \sqrt{\big(\sum_{g_3=1}^{G}\sum_{i_3,j_3\in\mathcal{T}_{g_3,n}}\hat{v}_{i_3,n}^2\big)\big(\sum_{g_3=1}^{G}\sum_{i_3,j_3\in\mathcal{T}_{g_3,n}}\hat{v}_{j_3,n}^2\big)}\\
&\leq \sqrt{\big(\mathcal{C}_{\mathcal{T},n}\sum_{k=1}^{n}\hat{v}_{k,n}^2\big)\big(\mathcal{C}_{\mathcal{T},n}\sum_{l=1}^{n}\hat{v}_{l,n}^2\big)}\\
&=\mathcal{C}_{\mathcal{T},n}(\sum_{i=1}^{n}\hat{v}_{i,n}^2).
\end{align*}

As a consequence,

\begin{displaymath}
\mathbb{V}[\frac{1}{n}\sum_{1\leq i \leq G_n}c_{ii,n}|\mathcal{X}_n, \mathcal{W}_n]=\frac{1}{n^2}\sum_{1\leq i \leq G_n}\mathbb{V}[c_{ii,n}|\mathcal{X}_n, \mathcal{W}_n]\leq\frac{1}{n^2}\sum_{1\leq i, j \leq G_n}\mathbb{V}[c_{ij,n}|\mathcal{X}_n, \mathcal{W}_n]= o_p(1),
\end{displaymath}
and
\begin{displaymath}
\mathbb{V}[\frac{1}{n}\sum_{1\leq i,j \leq G_n, i<j}c_{ij,n}|\mathcal{X}_n, \mathcal{W}_n]=\frac{1}{n^2}\sum_{1\leq i,j \leq G_n, i<j}\mathbb{V}[c_{ij,n}|\mathcal{X}_n, \mathcal{W}_n]\leq\frac{1}{n^2}\sum_{1\leq i, j \leq G_n}\mathbb{V}[c_{ij,n}|\mathcal{X}_n, \mathcal{W}_n]=o_p(1).\\
\end{displaymath}\\

In particular, $\tilde{\Sigma}_n(\bm{\kappa}_n)=\mathbb{E}[\tilde{\Sigma}_n(\bm{\kappa}_n)|\mathcal{X}_n, \mathcal{W}_n] + o_p(1)$, where
\begin{align*}
|\mathbb{E}[\tilde{\Sigma}_n(\bm{\kappa}_n)|\mathcal{X}_n, \mathcal{W}_n]| & \leq \frac{1}{n}\sum_{(g_1,i_1,j_1)}\sum_{(g_2,i_2,j_2)}\sum_{(g_3,i_3,j_3)}|\kappa_{g_1, g_2, i_1,j_1,i_2,j_2,n}||\hat{v}_{i_1,n}||\hat{v}_{j_1,n}||M_{i_2j_3,n}||M_{j_2i_3,n}||\mathbb{E}[U_{i_3,n}U_{j_3,n}|\mathcal{X}_n,\mathcal{W}_n]|\\
& \leq \mathcal{C}_{U,n} \frac{1}{n}\sum_{(g_1,i_1,j_1)}\sum_{(g_2,i_2,j_2)}\sum_{(g_3,i_3,j_3)}|\kappa_{g_1, g_2, i_1,j_1,i_2,j_2,n}||\hat{v}_{i_1,n}||\hat{v}_{j_1,n}||M_{i_2j_3,n}||M_{j_2i_3,n}|\\
& \leq \mathcal{C}_{U,n} \mathcal{C}_{\mathcal{T},n} \frac{1}{n}\sum_{(g_1,i_1,j_1)}\sum_{(g_2,i_2,j_2)}|\kappa_{g_1, g_2, i_1,j_1,i_2,j_2,n}||\hat{v}_{i_1,n}||\hat{v}_{j_1,n}|\\
& \leq \mathcal{C}_{U,n} \mathcal{C}_{\mathcal{T},n} \norm{\kappa_{n}}_{\infty} \frac{1}{n}\sum_{(g_1,i_1,j_1)}|\hat{v}_{i_1,n}||\hat{v}_{j_1,n}|\\
& \leq \mathcal{C}_{U,n} \mathcal{C}_{\mathcal{T},n}^2 \norm{\kappa_{n}}_{\infty} (\frac{1}{n}\sum_{1\leq i\leq n}\hat{v}_{i,n}^2) = O_p(1).
\end{align*}
We have therefore established that $\tilde{\Sigma}_n(\bm{\kappa}_n)=\mathbb{E}[\tilde{\Sigma}_n(\bm{\kappa}_n)|\mathcal{X}_n, \mathcal{W}_n] + o_p(1)$. It remains to show that $\hat{\Sigma}_n(\bm{\kappa}_n)=\tilde{\Sigma}_n(\bm{\kappa}_n)+o_p(1)$.\\
By using that $\hat{u}_{i,n}-\tilde{U}_{i,n}=\tilde{R}_{i,n}-\hat{v}_{i,n}(\hat{\beta}_n-\beta)$, where $\tilde{R}_{i,n}=\sum_{j=1}^{n}M_{ij,n}R_{j,n}$, we obtain
\begin{flalign*}
&\hat{\Sigma}_n(\bm{\kappa}_n)-\tilde{\Sigma}_n(\bm{\kappa}_n)=\frac{1}{n}\sum_{(g_1,i_1,j_1)}\sum_{(g_2,i_2,j_2)}\kappa_{g_1, g_2, i_1,j_1,i_2,j_2,n}\hat{v}_{i_1,n}\hat{v}_{j_1,n}(\hat{u}_{i_2,n}\hat{u}_{j_2,n}-\tilde{U}_{i_2,n}\tilde{U}_{j_2,n})\\
&=\frac{1}{n}\sum_{(g_1,i_1,j_1)}\sum_{(g_2,i_2,j_2)}\\
&\kappa_{g_1, g_2, i_1,j_1,i_2,j_2,n}\hat{v}_{i_1,n}\hat{v}_{j_1,n}[(\tilde{R}_{i_2,n} -\hat{v}_{i_2,n}(\hat{\beta}_n-\beta)+\tilde{U}_{i_2,n})(\tilde{R}_{j_2,n} -\hat{v}_{j_2,n}(\hat{\beta}_n-\beta)+\tilde{U}_{j_2,n})-\tilde{U}_{i_2,n}\tilde{U}_{j_2,n}].
\end{flalign*}
By the Cauchy-Schwarz inequality, it suffices to show that
\begin{align*}
&\frac{1}{n}\sum_{(g_1,i_1,j_1)}\sum_{(g_2,i_2,j_2)}\kappa_{g_1, g_2, i_1,j_1,i_2,j_2,n}\hat{v}_{i_1,n}^2(\tilde{R}_{i_2,n} -\hat{v}_{i_2,n}(\hat{\beta}_n-\beta))^2=o_p(1),\\
&\frac{1}{n}\sum_{(g_1,i_1,j_1)}\sum_{(g_2,i_2,j_2)}\kappa_{g_1, g_2, i_1,j_1,i_2,j_2,n}\hat{v}_{i_1,n}^2\tilde{U}_{j_2,n}^2=O_p(1).
\end{align*}
The latter can be straightforwardly shown by means of the arguments previously used to show $\tilde{\Sigma}_n(\bm{\kappa}_n)=O_p(1)$. For the former, since $\hat{v}_{j,n}=\tilde{V}_{j,n}+\tilde{Q}_{j,n}$, where $\tilde{Q}_{i,n}=\sum_{j=1}^{n}M_{ij,n}Q_{j,n}$, and $\tilde{R}_{i,n}=\tilde{r}_{i,n}+(\tilde{R}_{i,n}-\tilde{r}_{i,n})$, where $\tilde{r}_{i,n}=\sum_{j=1}^nM_{ij,n}r_{j,n}$ it suffices to show that
\begin{align*}
\frac{1}{n}\sum_{(g_1,i_1,j_1)}\sum_{(g_2,i_2,j_2)}|\kappa_{g_1, g_2, i_1,j_1,i_2,j_2,n}|\tilde{Q}_{i_1,n}^2\tilde{R}_{i_2,n}^2&=o_p(1),\\
\frac{1}{n}\sum_{(g_1,i_1,j_1)}\sum_{(g_2,i_2,j_2)}|\kappa_{g_1, g_2, i_1,j_1,i_2,j_2,n}|\tilde{V}_{i_1,n}^2\tilde{r}_{i_2,n}^2&=o_p(1),\\
\frac{1}{n}\sum_{(g_1,i_1,j_1)}\sum_{(g_2,i_2,j_2)}|\kappa_{g_1, g_2, i_1,j_1,i_2,j_2,n}|\tilde{V}_{i_1,n}^2|\tilde{R}_{i_2,n}-\tilde{r}_{i_2,n}|^2&=o_p(1),\\
(\hat{\beta}_n-\beta)^2\frac{1}{n}\sum_{(g_1,i_1,j_1)}\sum_{(g_2,i_2,j_2)}|\kappa_{g_1, g_2, i_1,j_1,i_2,j_2,n}|\hat{v}_{i_1,n}^2\hat{v}_{i_2,n}^2&=o_p(1).
\end{align*}
\par
First, $n^{-1}\sum_{(g_1,i_1,j_1)}\sum_{(g_2,i_2,j_2)}|\kappa_{g_1, g_2, i_1,j_1,i_2,j_2,n}|\tilde{V}_{i_1,n}^2\tilde{r}_{i_2,n}^2=o_p(1)$ because
\begin{align*}
\mathbb{E}\big[\frac{1}{n}\sum_{(g_1,i_1,j_1)}\sum_{(g_2,i_2,j_2)}|\kappa_{g_1, g_2, i_1,j_1,i_2,j_2,n}|\tilde{V}_{i_1,n}^2\tilde{r}_{i_2,n}^2|\mathcal{W}_n\big]&=\frac{1}{n}\sum_{(g_2,i_2,j_2)}\tilde{r}_{i_2,n}^2\sum_{(g_1,i_1,j_1)}|\kappa_{g_1, g_2, i_1,j_1,i_2,j_2,n}|\mathbb{E}\big[\tilde{V}_{i_1,n}^2|\mathcal{W}_n\big]\\
&\leq \mathcal{C}_{V,n} \mathcal{C}_{\mathcal{T},n}\frac{1}{n}\sum_{(g_2,i_2,j_2)}\tilde{r}_{i_2,n}^2\sum_{(g_1,i_1,j_1)}|\kappa_{g_1, g_2, i_1,j_1,i_2,j_2,n}|\\
&\leq \mathcal{C}_{V,n} \mathcal{C}_{\mathcal{T},n}\norm{\bm{\kappa}_{n}}_\infty(\frac{1}{n}\sum_{(g_2,i_2,j_2)}\tilde{r}_{i_2,n}^2)\\
&\leq \mathcal{C}_{V,n} \mathcal{C}_{\mathcal{T},n}^2\norm{\bm{\kappa}_{n}}_\infty(\frac{1}{n}\sum_{i=1}^n\tilde{r}_{i,n}^2)=O_p(\rho_n)=o_p(1),
\end{align*}
where the first inequality uses the fact that $\mathbb{E}[\tilde{V}_{i,n}|\mathcal{W}_n]\leq\mathcal{C}_{\mathcal{T},n}\mathcal{C}_{V,n}$, with $\mathcal{C}_{V,n}=1+\max_{1 \leq i \leq n},\mathbb{E}[\norm{V_{i,n}}^4|\mathcal{W}_n]$ as shown in \citeauthor{cjn} (\citeyear{cjn}, Supplemental Appendix).\par
Next,
\begin{align*}
\frac{1}{n}\sum_{(g_1,i_1,j_1)}\sum_{(g_2,i_2,j_2)}|\kappa_{g_1, g_2, i_1,j_1,i_2,j_2,n}|\tilde{V}_{i_1,n}^2|\tilde{R}_{i_2,n}-\tilde{r}_{i_2,n}|^2&\leq n \norm{\bm{\kappa}_{n}}_\infty(\frac{1}{n}\sum_{(g_1,i_1,j_1)}\tilde{V}_{i_1,n}^2)(\frac{1}{n}\sum_{(g_2,i_2,j_2)}|\tilde{R}_{i_2,n}-\tilde{r}_{i_2,n}|^2)\\
&\leq n \norm{\bm{\kappa}_{n}}_\infty\mathcal{C}_{\mathcal{T},n}^2(\frac{1}{n}\sum_{i=1}^n\tilde{V}_{i,n}^2)(\frac{1}{n}\sum_{i=1}^n|\tilde{R}_{i,n}-\tilde{r}_{i,n}|^2)\\
&=O_p[n(\varrho_n-\rho_n)]=o_p(1)
\end{align*}
and
\begin{align*}
\frac{1}{n}\sum_{(g_1,i_1,j_1)}\sum_{(g_2,i_2,j_2)}|\kappa_{g_1, g_2, i_1,j_1,i_2,j_2,n}|\tilde{Q}_{i_1,n}^2\tilde{R}_{i_2,n}^2&\leq n \norm{\bm{\kappa}_{n}}_\infty(\frac{1}{n}\sum_{(g_1,i_1,j_1)}\tilde{Q}_{i_1,n}^2)(\frac{1}{n}\sum_{(g_2,i_2,j_2)}\tilde{R}_{i_2,n}^2)\\
&\leq n \norm{\bm{\kappa}_{n}}_\infty\mathcal{C}_{\mathcal{T},n}^2(\frac{1}{n}\sum_{i=1}^n\tilde{Q}_{i,n}^2)(\frac{1}{n}\sum_{i=1}^n\tilde{R}_{i,n}^2)\\
&=O_p(n\chi_n\varrho_n)]=o_p(1)
\end{align*}\par
Finally,
\begin{align*}
(\hat{\beta}_n-\beta)^2\frac{1}{n}\sum_{(g_1,i_1,j_1)}\sum_{(g_2,i_2,j_2)}|\kappa_{g_1, g_2, i_1,j_1,i_2,j_2,n}|\hat{v}_{i_1,n}^2\hat{v}_{i_2,n}^2=o_p(1)
\end{align*}
because $\sqrt[]{n}(\hat{\beta}_n-\beta)=O_p(1)$ and
\begin{align*}
\frac{1}{n^2}\sum_{(g_1,i_1,j_1)}\sum_{(g_2,i_2,j_2)}|\kappa_{g_1, g_2, i_1,j_1,i_2,j_2,n}|\hat{v}_{i_1,n}^2\hat{v}_{i_2,n}^2&\leq (\max_{1\leq i\leq n} |\hat{v}_{i,n}|)^2\frac{1}{n^2}\sum_{(g_1,i_1,j_1)}\sum_{(g_2,i_2,j_2)}|\kappa_{g_1, g_2, i_1,j_1,i_2,j_2,n}|\hat{v}_{i_2,n}^2\\
&\leq (\frac{\max_{1\leq i\leq n} |\hat{v}_{i,n}|}{\sqrt{n}})^2\norm{\bm{\kappa}_{n}}_\infty(\frac{1}{n}\sum_{(g_2,i_2,j_2)}\hat{v}_{i_2,n}^2)\\
&\leq (\frac{\max_{1\leq i\leq n} |\hat{v}_{i,n}|}{\sqrt{n}})^2\norm{\bm{\kappa}_{n}}_\infty\mathcal{C}_{\mathcal{T},n}(\frac{1}{n}\sum_{i=1}^n\hat{v}_{i,n}^2)\\
&=o_p(1),
\end{align*}
which concludes the proof.

\subsection{Proof of Lemma 2}
Let us define $d_{i_1j_1,i_3j_3,n}=\sum_{(g_2, i_2, j_2)}\kappa_{g_1,g_2,i_1,j_1,i_2,j_2,n}M_{i_3,j_2,n}M_{j_3,i_2,n}-\mathbbm{1}\{(i_1,j_1)=(i_3,j_3)\}$. We hence have
\begin{equation*}
\mathbb{E}[\tilde{\Sigma}_n(\bm{\kappa}_n)|\mathcal{X}_n, \mathcal{W}_n]-\Sigma_n= \frac{1}{n}\sum_{(g_1,i_1,j_1)}\sum_{(g_3,i_3,j_3)}d_{i_1j_1,i_3j_3,n}\hat{v}_{i_1,n}\hat{v}_{j_1,n}\mathbb{E}[U_{i_3,n}U_{j_3,n}|\mathcal{X}_n, \mathcal{W}_n],
\end{equation*}
so if $\max_{(g_1,i_1,j_1)}\sum_{(g_3,i_3,j_3)}|d_{i_1j_1,i_3j_3,n}|=o_p(1)$, then
\begin{align*}
|\mathbb{E}[\tilde{\Sigma}_n(\bm{\kappa}_n)|\mathcal{X}_n, \mathcal{W}_n]-\Sigma_n|&\leq \frac{1}{n}\sum_{(g_1,i_1,j_1)}\sum_{(g_3,i_3,j_3)}|d_{i_1j_1,i_3j_3,n}||\hat{v}_{i_1,n}||\hat{v}_{j_1,n}||\mathbb{E}[U_{i_3,n}U_{j_3,n}|\mathcal{X}_n, \mathcal{W}_n]|\\
&\leq \mathcal{C}_{U,n}\frac{1}{n}\sum_{(g_1,i_1,j_1)}\sum_{(g_3,i_3,j_3)}|d_{i_1j_1,i_3j_3,n}||\hat{v}_{i_1,n}||\hat{v}_{j_1,n}|\\
&\leq \mathcal{C}_{U,n}(\frac{1}{n}\sum_{(g_1,i_1,j_1)}|\hat{v}_{i_1,n}||\hat{v}_{j_1,n}|)\big(\max_{(g_1,i_1,j_1)}\sum_{(g_3,i_3,j_3)}|d_{i_1j_1,i_3j_3,n}|\big)\\
&\leq \mathcal{C}_{U,n}\mathcal{C}_{\mathcal{T},n}(\frac{1}{n}\sum_{i=1}^{n}\hat{v}_{i,n}^2)\big(\max_{(g_1,i_1,j_1)}\sum_{(g_3,i_3,j_3)}|d_{i_1j_1,i_3j_3,n}|\big)=o_p(1).
\end{align*}

\subsection{Proof of Lemma 3}
If $\lambda_{\text{min}}(\mathbf{S}_n'(\mathbf{M}_n\otimes\mathbf{M}_n) \mathbf{S}_n)>0$, then
\begin{equation*}
\Big| \sum_{(g_2, i_2, j_2)}\kappa_{g_1,g_2,i_1,j_1,i_2,j_2,n}^{\texttt{CR}}M_{i_1j_2,n}M_{j_1i_2,n}-1\Big|+\sum_{(g_3,i_3,j_3)\neq(g_1,i_1,j_1)}\Big|\sum_{(g_2, i_2, j_2)}\kappa_{g_1,g_2,i_1,j_1,i_2,j_2}^{\texttt{CR}}M_{i_3j_2,n}M_{j_3i_2,n}\Big|= 0,
\end{equation*}
which combined with Lemma 2 gives $\mathbb{E}[\bm{\tilde{\Sigma}}_n(\bm{\kappa}_n^{\textup{\texttt{CR}}})|\mathcal{X}_n, \mathcal{W}_n]=\bm{\Sigma}_n + o_p(1)$.

\subsection{Proof of Lemma 4}
Recall that for LZ's estimator we have
\begin{align*}
&\sum_{g_2,i_2,j_2}|d_{i_1,j_1,i_2,j_2,n}|=|M_{i_1i_1,n}M_{j_1j_1,n}-1|+\sum_{(g_2,i_2,j_2)\neq (g_1,i_1,j_1)}|M_{i_1j_2,n}||M_{j_1i_2,n}|\\
&=(1-M_{i_1i_1,n}M_{j_1j_1,n})+M_{i_1i_1,n}(\sum_{\substack{i_2\in g_1\\i_2\neq j_1}}|M_{j_1i_2,n}|)+M_{j_1j_1,n}(\sum_{\substack{j_2\in g_1\\j_2\neq i_1}}|M_{i_1j_2,n}|)+\sum_{\substack{(g_2,i_2,j_2)\\ i_2\neq j_1, j_2\neq i_1}}|M_{i_1,j_2,n}||M_{j_1,i_2,n}|.
\end{align*}
Defining $\mathcal{M}_n=1-\min_{1\leq i\leq n}M_{ii,n}$, we have that $\mathcal{M}_n=O_p(\frac{K_n}{n})$ (see \citealp{cjn}) and
\begin{equation*}
\max_{\substack{i,j\\i\neq j}}|M_{ij,n}|\leq\max_{\substack{i,j\\i\neq j}}\frac{1}{n}\sum_{l=1}^{K_n}|w^{*}_{il,n}||w^{*}_{jl,n}|\leq\max_{i}\frac{1}{n}\sum_{l=1}^{K_n}w_{il,n}^{*2}=O_p(\frac{K_n}{n}).
\end{equation*}
As a result, we have
\begin{align*}
\max_{(g_1,i_1,j_1)}\sum_{g_2,i_2,j_2}|d_{i_1,j_1,i_2,j_2,n}| &\leq 2\mathcal{M}_n +2(\mathcal{C}_{\mathcal{T},n}-1)O_p(\frac{K_n}{n})+\mathcal{C}_{\mathcal{T},n}^2G_{n}O_p(\frac{K_n^2}{n^2})\\
&\leq O_p(\frac{K_n}{n})+2(\mathcal{C}_{\mathcal{T},n}-1)O_p(\frac{K_n}{n})+\mathcal{C}_{\mathcal{T},n}^2O(n)O_p(\frac{K_n^2}{n^2})= O_p(\frac{K_n^2}{n}),
\end{align*}
which combined with Lemma 2 gives $\mathbb{E}[\tilde{\Sigma}_n(\mathbf{I}_{L_n})|\mathcal{X}_n, \mathcal{W}_n]=\Sigma_n + o_p(1)$.

\subsection{Proof of Lemma 5}
Under homoskedasticity one has
\begin{align*}
\mathbb{E}[\tilde{\Sigma}_n(\mathbf{I}_{L_n})|\mathcal{X}_n, \mathcal{W}_n]&=\frac{\sigma^2_n}{n}\sum_{(g_1,i_1,j_1)}\sum_{k=1}^n\hat{v}_{i_1,n}\hat{v}_{j_1,n}M_{i_1k,n}M_{j_1k,n}\\
&=\frac{\sigma^2_n}{n}\sum_{(g_1,i_1,j_1)}\hat{v}_{i_1,n}\hat{v}_{j_1,n}M_{i_1j_1,n},
\end{align*}
and
\begin{equation*}
\Sigma_n=\frac{\sigma^2_n}{n}\sum_{i=1}^n\hat{v}^2_{i,n}.
\end{equation*}
As a result, we have
\begin{align*}
|\mathbb{E}[\tilde{\Sigma}_n(\mathbf{I}_{L_n})|\mathcal{X}_n, \mathcal{W}_n]-\Sigma_n|&\leq \frac{\sigma^2_n}{n}\sum_{i=1}^{n}\hat{v}^2_{i,n}|M_{ii.n}-1| + \frac{\sigma^2_n}{n}
\sum_{\substack{(g_1,i_1,j_1) \\i_1\neq j_1}}|\hat{v}_{i_1,n}||\hat{v}_{j_1,n}||M_{i_1j_1,n}|\\
&\leq \mathcal{C}_{U,n}\mathcal{M}_n(\frac{1}{n}\sum_{i=1}^n\hat{v}_{i,n}^2) + \mathcal{C}_{U,n}\mathcal{C}_{\mathcal{T},n}(\max_{\substack{i,j\\i\neq j}}|M_{ij,n}|)(\frac{1}{n}\sum_{i=1}^n\hat{v}_{i,n}^2)\\
&\leq \mathcal{C}_{U,n}O_p(\frac{K_n}{n})(\frac{1}{n}\sum_{i=1}^n\hat{v}_{i,n}^2)+  \mathcal{C}_{U,n}\mathcal{C}_{\mathcal{T},n}O_p(\frac{K_n}{n})(\frac{1}{n}\sum_{i=1}^n\hat{v}_{i,n}^2)=O_p(\frac{K_n}{n}),
\end{align*}
and therefore $\mathbb{E}[\tilde{\Sigma}_n(\mathbf{I}_{L_n})|\mathcal{X}_n, \mathcal{W}_n]= \Sigma_n +o_p(1)$.

\section{Extension to within-cluster restrictions}\label{sectionF}
In this section, we present an extension of the class of estimators studied in this paper that allows to impose zero restrictions on the variance-covariance matrix of the errors within clusters.

Define the sets
\begin{align*}
&\mathcal{V}_{g,n}=\{(i,j)\in \mathcal{T}_{g,n}\times\mathcal{T}_{g,n}: \mathbb{E}[U_{i,n}U_{j,n}|\mathcal{X}_n, \mathcal{W}_n]\neq0\},\\
&\mathcal{R}_{g,i,n}=\{j\in \mathcal{T}_{g,n}: \mathbb{E}[U_{i,n}U_{j,n}|\mathcal{X}_n, \mathcal{W}_n]\neq0\},
\end{align*}
and let $L_n$ be the number of non-zero elements contained in $\bm{\Omega}_{U,n}=\mathbb{E}[\mathbf{U}_n\mathbf{U}_n'|\mathcal{X}_n,\mathcal{W}_n]$. The generalized version of our proposed class of cluster-robust variance estimators reads:
\begin{equation*}
\bm{\hat{\Sigma}}_n^{}(\bm{\kappa}_n)=\frac{1}{n}\sum_{g_1=1}^{G_n}\sum_{g_2=1}^{G_n}\sum_{(i_1,j_1)\in\mathcal{V}_{g_1,n}}\sum_{(i_2,j_2)\in\mathcal{V}_{g_2,n}}\kappa_{g_1, g_2, i_1,j_1,i_2,j_2,n}\mathbf{\hat{v}}_{i_1,n}\mathbf{\hat{v}}'_{j_1,n}\hat{u}_{i_2,n}\hat{u}_{j_2,n},
\end{equation*}
where $\kappa_{g_1,g_2,i_1,j_1,i_2.j_2,n}$ corresponds to the $(h(g_1,i_1,j_1), h(g_2,i_2,j_2))$ entry of the $L_n\times L_n$ symmetric matrix $\bm{\kappa}_{n}$, where $h(g,i,j)=[\sum_{k=0}^{(g-1)}(\#\mathcal{V}_{k,n})+\sum_{k=0}^{i-1}(\#\mathcal{R}_{g,k,n})+\overline{j(i)_{g,n}}]$ with $\overline{j(i)_{g,n}}=\#\{k\in \mathcal{T}_{g,n}: \mathbb{E}[U_{i,n}U_{j,n}|\mathcal{X}_n, \mathcal{W}_n]\neq0 \quad \text{and}\quad k\leq j\}$ and we adopt the convention that $\#\mathcal{V}_{0,n}=0$ and $\#\mathcal{R}_{g,0,n}=0 \quad \forall g$.

A consistent estimator under Assumptions 1*-3* is then defined as $\bm{\hat{\Sigma}}(\bm{\kappa}_n^{\texttt{CR}})$, where $\bm{\kappa}^{\texttt{CR}}_n=(\mathbf{S}_n'(\mathbf{M}_n\otimes\mathbf{M}_n) \mathbf{S}_n)^{-1}$.

\end{document}